\documentclass[journal,12pt,onecolumn]{IEEEtran}
\usepackage[margin=1 in]{geometry}
\normalsize
\usepackage[bottom]{footmisc}
\usepackage{color}
\usepackage{graphicx}
\usepackage{amsfonts}
\usepackage[cmex10]{amsmath}
\usepackage{cite}
\usepackage{enumitem}
\usepackage[center]{caption}
\usepackage{epsfig}
\usepackage{latexsym}
\usepackage{amsthm}
\usepackage{bbm}
\usepackage{chngcntr}

\usepackage{mathtools}
\graphicspath{{./figures/}}
\theoremstyle{definition}
\newtheorem{definition}{Definition}
\newtheorem{example}{Example}

\newtheorem{construction}{Construction}
\newtheorem{theorem}{Theorem}
\newtheorem{corollary}{Corollary}
\newtheorem{remark}{Remark}
\newtheorem{lemma}{Lemma}
\newcounter{numrel}
\counterwithin*{numrel}{section}

\newcommand{\remove}[1]{}

\usepackage{amssymb}
\DeclarePairedDelimiter\ceil{\lceil}{\rceil}

\DeclareOldFontCommand{\rm}{\normalfont\rmfamily}{\mathrm}
\begin{document}
	\title{Fundamental Limits of Erasure-Coded Key-Value Stores with Side Information}
	\author{Ramy E. Ali, Viveck R. Cadambe, Jaime Llorca and Antonia M. Tulino}
\maketitle
\begin{abstract}
 In applications of distributed storage systems to modern key-value stores, the stored data is highly dynamic due to frequent updates. The multi-version coding problem was formulated to study the cost of storing dynamic data in distributed storage systems. Previous work on multi-version coding considered a completely decentralized and asynchronous system assuming that the servers are not aware of which versions of the data are received by the other servers. In this paper, we relax this assumption and study a system where a server may acquire side information of the data versions propagated to some other servers based on the network topology. Specifically, we study a storage system with $n$ servers over a directed graph that store $\nu$ totally ordered versions of a message. Each server receives a subset of these $\nu$ versions. A server is aware of which versions have been received by its neighbors in the network graph. We show that the side information can result in a better storage cost as compared with the case where there is no side information for some regimes at the expense of the additional latency associated with exchanging the side information. Through an information-theoretic converse, we identify surprising scenarios where the side information may not help in improving the worst-case storage cost beyond the case where servers have no side information. Finally, we present a case study over Amazon web services (AWS) that demonstrates the potential cost reductions that may be obtained by our constructions.
\end{abstract}

\IEEEpeerreviewmaketitle
\section{Introduction}
\makeatletter{\renewcommand*{\@makefnmark}{}
	\footnotetext{\hrule \vspace{0.05in} Ramy E. Ali (E-mail: ramy.ali@psu.edu) is with the School of Electrical Engineering and Computer Science, The Pennsylvania State University, University Park, PA and was with Nokia Bell Labs, Holmdel, NJ. Viveck R. Cadambe (E-mail: viveck@engr.psu.edu) is with the School of Electrical Engineering and Computer Science, The Pennsylvania State University, University Park. Jaime Llorca (E-mail: jaime.llorca@nokia.com) is with Nokia Bell Labs, Holmdel, NJ. Antonia M. Tulino (E-mail: a.tulino@nokia.com or antoniamaria.tulino@unina.it) is with Nokia Bell Labs, Holmdel, NJ and with the Department of Electrical Engineering and Information Technology, University of Naples Federico II, Naples, Italy. This work is supported by NSF grant No. CCF 1553248 and is published in part in the proceedings of the 2018 IEEE International Symposium on Information Theory, Vail, Colorado \cite{ali2018multi}.}

Distributed key-value stores such as Apache Cassandra \cite{CassandradB} and Amazon Dynamo DB \cite{Decandia} form an integral part of modern cloud computing infrastructure. Key-value stores are used by several applications such as reservation systems, transactions and multi-player gaming. Distributed key-value stores are engineered to deliver data with low latencies, as well as to minimize costs for the provider incurred in terms of memory and communication. Such key-value stores replicate data to ensure availability of data for a wide geographic area as well as for fault tolerance. Much research has considered reducing costs and improving latency of geo-distributed key-value stores, including various consistency models, memory management policies, and optimal data placement (See \cite{Decandia, CassandradB, vogels2008eventually, bailis2013eventual, Causal_Consistency1, Causal_Consistency2, Causal_Consistency3, shen2015causal, spanstore} and references therein). 

In this paper we focus on the use of \emph{erasure coding} for reducing costs of geo-distributed key-value stores by studying such systems from an information-theoretic perspective. Erasure coding is a generalization of replication that is well known to incur much lower storage costs for the same degree of fault tolerance. Although erasure coding comes at the cost of increased computational complexity for writing (encoding) and reading (decoding) data, the potential cost-savings in the face of rapidly growing data volumes has made erasure coding increasingly attractive \cite{ intel-isal, rashmi2016ec}. Indeed, recent research in erasure coding for distributed storage has been marked by some remarkable advances in coding theory \cite{Tamo_Barg, Dimakis, Cheng_Erasurecoding, gopalan2012locality} and their implementation for \emph{archival}$^\mathrm{1}$\footnote{$^\mathrm{1}$Archival storage systems are whose where the data does not change frequently.} storage systems \cite{huang2012erasure,FacebookF4}. 
 \begin{figure}[h]
\centering
\includegraphics[width=.6\textwidth,height=.25\textheight]{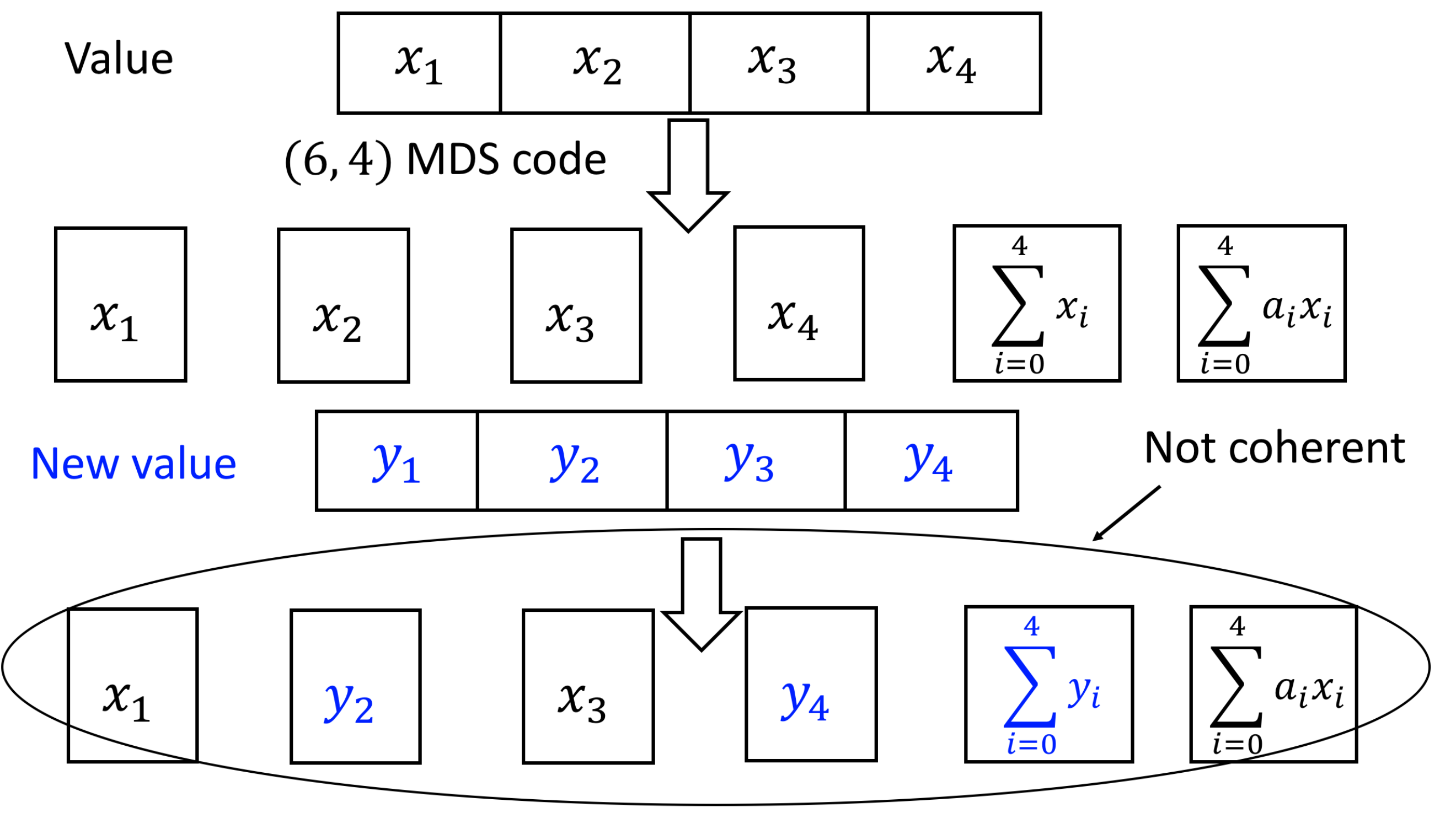}
\caption{\footnotesize{A value $(x_1, x_2, x_3, x_4)$} is encoded using $(6, 4)$ maximum distance separable (MDS) code and each node stores a codeword symbol. The message is then updated to $(y_1, y_2, y_3, y_4)$, but only three nodes store the updated codeword symbols and the other three nodes store the old codeword symbols.}
\label{fig:update_coding}
\end{figure}

While much research has studied erasure coding for archival storage in recent years, the study of erasure coding specifically tailored for key-value stores, which are non-archival storage systems modeled over decentralized asynchronous networks, is in its infancy. In fact, the use of erasure codes in key-value stores face some unique challenges that do not appear in archival storage. The recent works \cite{Cheng_Erasurecoding, Cachin_Tessaro, Dutta, Hendricks, GWGR, cadambe2017coded, Konwar_ipdps2016, konwar_opodis2016, dobre_powerstore, AWE } developed protocols that modify replication-based protocols and use erasure codes. Recall that an $(n, k)$ code - where $n$ represents the code length and $k$ represents the dimension of the code - partitions the value to be stored into $k$ data fragments and encodes them into $n$ coded fragments such that each storage node stores one coded fragment. The use of maximum distance separable (MDS) codes (e.g. Reed-Solomon codes) ensures that the value can be recovered from any $k$ of the $n$ nodes. We notice that for $k > 1$, there is no single node that stores the data entirely. \\ 
When such a code is used in a key-value store, when a \emph{write} operation updates the data, a node has to effectively wait for a sufficient number of nodes to receive the update before deleting the old version of the data (See Fig. \ref{fig:update_coding}). There are two approaches to solve this problem. A common approach is to allow nodes to store older versions of data, so that in the duration of the propagation of the new version, there is indeed a coherent version of the data that can be decoded. In fact, it is demonstrated that for such key-value stores algorithms \cite{Hendricks, cadambe2017coded, Dutta, GWGR, AWE}, each node has to store a number of versions that is linear in the degree of concurrency - the number of concurrent writes to the same object. However, the need to store older versions offsets some of the storage cost benefits of erasure coding. A second approach used by protocols \cite{Cachin_Tessaro, Konwar_ipdps2016} gets around the worst case storage growth with concurrency through significantly higher communication. These protocols use message broadcast primitives where the nodes exchange data among themselves to find out the versions received by the other nodes and delete the older versions. Although these protocols store only one version of the data object per server, they can incur a significant latency  especially in geo-distributed settings, as the clients have to wait for the information to propagate to large number of geographically spread out data centers before returning. Furthermore, this information exchange comes at significant communication costs, as inter-data center communication can be quite expensive. 
 
 In this paper, we depart from the previous approaches through the development of new erasure codes specifically tailored for geo-distributed key-value stores by studying them from an information-theoretic perspective. In particular, we show that conventional approaches that use standard erasure coding can be too pessimistic, and they miss opportunities to reduce the storage costs. Our coding scheme exploits the inherent network topology of the nodes storing an object, and demonstrates that exchanging some small amount of gossip information can reduce the costs incurred in erasure coding based key-value stores. Our approach can be viewed as a bridge between \cite{cadambe2017coded, Hendricks, Dutta, dobre_powerstore}, where no side information is exchanged among the nodes, and \cite{Cachin_Tessaro, Konwar_ipdps2016} where the broadcast primitives are used to exchange complete side information among the nodes of the system$^\mathrm{2}$.  \footnote
 {$^\mathrm{2}$We note that we mainly focus on the coding aspects and do not design an explicit protocol.}
 
 We describe the central technical contribution of our work through the \emph{multi-version coding} framework \cite{wang2018multi}. Specifically, this framework abstracts out algorithmic details of key-value stores to focus on the coding-related details while retaining the essence of consistent storage systems. It thereby provides a method of developing codes for key-value stores and studying their storage costs. Achievable coding schemes for multi-version coding have been used in protocols for key-value stores in \cite{Konwar_TREASMOD} and information-theoretic converses for multi-version coding have led to impossibility results that bound the costs incurred by such protocols \cite{Cadambe_Wang_Lynch2016}. In contrast with \cite{wang2018multi}, which pessimistically assumes no side information - that is that each storage node does not know  the versions received by other nodes - we assume the storage nodes can exchange meta-data information of versions received with nearby nodes, especially if the data exchange can be relatively inexpensive. The nodes that exchange this meta-data side information induce a \emph{side information graph}; the vertices of the side information are the storage nodes and an edge between two nodes implies that these nodes exchange meta-data. \\ We determine erasure coding strategies based on the topology of the side information graph and study their storage costs. One important technical contribution we make is the identification of a graph-theoretic functional - the size of the \emph{complement of the smallest maximally externally connected subset (CSMECS)} - that determines the costs of our coding strategies. We also develop information-theoretic impossibility results that bound the storage cost for a given topology. Although the central contributions of our paper are coding-theoretic in nature, we complement our results with a case study where we consider a hypothetical geo-distributed key-value store over Amazon web services (AWS). By deriving a side information graph based on the communication costs and latencies between data centers, and deriving the implied (projected) storage costs, we showcase the potential impact of our results.  
\section{Background and Summary of Contributions}

We begin with a background of quorum-based algorithms for key-value stores. We then describe the multi-version coding framework and summarize our main contributions. 
 \subsection{Quorum-based  Algorithms for  Consistent Key-Value Stores} 
We focus on key-value stores that offer simple read (get) and write (put) operations using on \emph{quorum-based} algorithms. Consider a setting with $n$ storage nodes. In quorum-based protocols that use replication, e.g. the  well-known ABD algorithm \cite{ABD}, a write operation sends a write request to all nodes and waits for the responses of at least $c_W$ nodes before completing the write operation, where $c_W, 1 \leq c_W \leq n$ is the write quorum size. Similarly, a read operation sends a read request to all $n$ nodes and waits for responses from at least $c_R$ nodes in the system before returning the value, where $c_R, 1 \leq c_R \leq n$ is the size of the read quorum.

 To ensure a failure tolerance of $f$ nodes, it is required that $c_W, c_R \leq n-f$. Key-value protocols such as \cite{ABD} order the different writes$^\mathrm{3}$\footnote{$^\mathrm{3}$\noindent These protocols usually use Lamport clocks to order the different writes; our formulation here abstracts out these details.}, and require that the latest version that has propagated to at least $c_W$ nodes - that is, the version corresponding to latest \emph{complete} write operation - can be obtained by a reader that connects to any $c_R$ nodes. Notice that for every pair of completed write and read operations, there are at least $c_R+c_W-n$ nodes that received the value of the write operation, and responded to the read operation. For a replication-based algorithm, $c_W$ and $c_R$ are chosen such that $c_W+c_R>n$ to ensure that the latest complete version can be obtained. The use of an MDS code with dimension $k$ would require a read operation to get data from at least $k$ nodes with the \emph{same version} of the object. Thus, $c_W$ and $c_R$ are chosen such that $c_W+c_R-n \geq k$ to ensure that latest complete version can be decoded.
\subsection{Multi-Version Coding}
The multi-version coding framework formalizes the storage strategies of key-value stores through an information-theoretic framework. Multi-version coding considers a distributed storage system of $n$ nodes that store $\nu$ totally ordered independent versions of a  $K$ bits object. The higher ordered versions are interpreted as later versions, and lower ordered versions as earlier versions. Due to the inherent asynchrony in the system, the versions may not propagate to all the nodes. Specifically, each node receives an arbitrary subset of these $\nu$ versions denoting the \emph{state} of that node. Any version that has been propagated to at least $c_W$ nodes is dubbed a \emph{complete} version, and the goal at the decoder is to connect to an arbitrary subset of $c_R$ nodes and decode the latest complete version - the complete version with the highest order - or a later version.

For any complete version and for any set of $c_R$ nodes, there are at least $c \coloneqq c_W+c_R-n$ nodes that have received that version. In the classical erasure coding model, where $\nu=1$, the Singleton bound implies that that the storage cost per node is at least $K/c$.  However for $\nu > 1$, a node cannot simply store the codeword symbol corresponding to one version, since other nodes may not have received that version. In fact, the lower bound of \cite{wang2018multi} implies that the amount of information to be stored is at least $\frac{\nu K}{c+\nu-1}-\theta(1) 
=(\frac{\nu}{c} - \frac{\nu(\nu-1)}{c^2}+o(\frac{1}{c^2})) K$. That is, there is a  cost to be paid for the decentralized nature and the asynchrony in the system, and this cost grows with $\nu$ which intuitively corresponds to the degree of concurrency in the system. The impact of correlations has been studied in multi-version coding setting in \cite{ali2017harnessing}. Like all models, multi-version coding does not perfectly capture all aspects of key-value stores, yet the insights it obtains are useful. In particular, although the framework abstracts out several algorithmic issues, insights of multi-version coding have been used to develop low cost key-value store protocols \cite{Zorgui_MVC} as well as impossibility results that bounds the cost of such protocols \cite{Cadambe_Wang_Lynch2016}. 

\subsection{Contributions}

In this paper, we extend the scope of the multi-version coding framework, motivated by protocols such as \cite{Cachin_Tessaro, Konwar_ipdps2016} where nodes exchange side information among themselves to reduce the storage cost. Our formulation is also motivated by real-world considerations of geo-distributed data stores where such data exchange can be more feasible between certain sets of nodes, depending on the topology of the nodes (data centers) and latency requirements. Specifically, we allow nodes to receive side information of the states of some other nodes based on the network topology and study the impact of this side information on the storage cost. We represent the side information by a directed graph $\mathcal G=(\mathcal N, \mathcal E)$, where an edge $e_{ij} \in \mathcal E$ from vertex $i$ to vertex $j$ indicates that node $i$ is aware of the state of node $j$.

In the completely centralized case where each node is aware of the states of all nodes, that is $\mathcal G=(\mathcal N, \mathcal E)$ is a complete digraph, each node is aware of the latest complete version. In this case, a node that receives the latest complete version stores it with an MDS code of dimension $c$. Therefore, the storage cost is $\frac{K}{c}$. In the completely decentralized setting studied in \cite{wang2018multi}, where $\mathcal G=(\mathcal N, \mathcal E)$ has no edges except the self edges, the storage cost is at least $(\frac{\nu}{c} - \frac{\nu(\nu-1)}{c^2}+o(\frac{1}{c^2})) K$. Here, we provide results that depart from the the two extreme points - the completely centralized setting where classical erasure coding-based bounds and constructions suffice, and the multi-version coding setting which is completely decentralized - and bridge the gap between them. Specifically, our contributions are as follows. 
\begin{enumerate}
	\item We provide code constructions that show that this side information can reduce the worst-case storage cost significantly as compared with the case where nodes do not share their states. In particular, for a given side information graph, we identify a graph functional, that we refer to as the size of the \emph{complement of the smallest $c_W$-maximally externally connected subset}- or simply, the size of the $c_W$-CSMECS - denoted by $\overline{m}_{\mathcal G}$, which dictates the storage cost of our code constructions. Specifically, our construction has a storage cost of $ \left( \frac{1}{c}+\frac{(\nu-1) \overline{m}_{\mathcal G}( c_W)}{c^2}+o\left(\frac{\overline{m}_{\mathcal G}( c_W)}{c^2} \right) \right)K$. In a fully connected topology, $\overline{m}_{\mathcal G}(c_W) = 0$, whereas for a completely decentralized setting, $\overline{m}_{\mathcal{G}}(c_W)$ is equal to $c_W-1$. For a regular side information graph $\mathcal G$ with degree $H$, we show that $\overline{m}_{\mathcal{G}}(c_W) \leq \min ((n-c_W+1)(n-H), c_W-1)$, which leads to a an achievable scheme on the storage cost.
	\item We also provide information-theoretic lower bounds for the case of $\nu=2$ and identify a curious outcome of these results. Specifically, we identify a scenario where each node is aware of the versions received by $(n-3)$ other nodes, and yet this tremendous amount of side information does not help in improving the worst-case storage cost. These results indicate that a careful understanding of the topology is required to completely exploit the side information in distributed key-value stores.
\end{enumerate}

\subsection{Case Study}
Several services implement their own key-value stores using public cloud services, for instance, Overleaf~\cite{overleaf} uses Amazon S3~\cite{s3}. The development of a full-fledged protocol that use our code constructions, and implementation of such a protocol is outside the scope of our work. However, we conduct a case study in Section \ref{Amazon Examples} of the potential cost savings of using our code constructions assuming a hypothetical key-value store implementation over Amazon web services public cloud. While our constructions would also be relevant for private/commercial key-value stores such as DynamoDB, the fact that the pricing information is readily available in the public cloud setting enables us to obtain a realistic understanding of the potential utility of our contributions. Current protocols either use replication \cite{ABD}, erasure coding that is completely decentralized \cite{CadambeCoded_NCA, Dutta, Hendricks, AWE, dobre_powerstore}, or assume full data exchange among the nodes and boil down to erasure coding solutions that are completely centralized \cite{Cachin_Tessaro, Konwar_ipdps2016}. Based on the latencies between data centers, we construct a side information graph $\mathcal{G}$ and calculate the size of the $c_W$-CSMECS $\overline m_{\mathcal{G}}(c_W)$ for the generated graph and show that using partial side information and our code constructions can lead to cost savings in comparison to current approaches.

\subsection*{Organization of the paper}
The rest of this paper is organized as follows. In Section \ref{Model}, we formulate the multi-version coding problem with side information. In Section \ref{Code Construction}, we provide our code constructions. In Section \ref{Impossibility Results}, we develop a lower bound on the per-server worst case storage cost. In Section \ref{Numerical Examples}, we provide numerical examples and a case study showing the storage gain of our code constructions. Finally, conclusions and future work are discussed in Section \ref{Conclusion}.

\section{System Model}
\label{Model}
We start with the notation. For a positive integer $i$, we denote by $[i]$ the set $\lbrace 1, 2, \cdots, i\rbrace$. For any set of ordered indices $S=\lbrace s_1, s_2, \cdots, s_{|S|}\rbrace \subseteq \mathbb{Z}$, where $s_1< s_2 < \cdots < s_{|S|}$, and for any ensemble of variables $\lbrace X_i : i \in S\rbrace$, the tuple $(X_{s_1}, X_{s_2}, \cdots, X_{s_{|S|}})$ is denoted by $X_S$. We use $\log (.)$ to denote the logarithm to the base $2$ and $H(.)$ to denote the binary entropy function. We use the notation $[2^K]$ to denote the set of $K$-length binary strings. A \emph{code} of length $n$ and dimension $k$ over alphabet $\mathcal{A}$ consists of an injective mapping $\mathcal{C}:\mathcal{A}^{k} \rightarrow \mathcal{A}^{n}$. When $\mathcal{A}$ is a finite field and the mapping $\mathcal{C}$ is linear, then the code is referred to as a \emph{linear code}. A linear code $\mathcal{C}$ of length $n$ and dimension $k$ is referred to as $(n,k)$ code. An $(n,k)$ linear code is referred to as maximum distance separable (MDS) if the mapping projected to \emph{any} $k$ co-ordinates is invertible. In a directed graph $\mathcal G=(\mathcal V, \mathcal E)$, the in-degree of a vertex $v \in \mathcal V$ is denoted by $\mathrm{deg}^{-}_{\mathcal G}(v)$ and the out-degree of $v$ is denoted by $\mathrm{deg}^{+}_{\mathcal G}(v)$.

We study a storage system of $n$ servers, denoted by $\mathcal N$, that can tolerate $f$ failures$^\mathrm{4}$\footnote{$^\mathrm{4}$By failures, we refer to servers that halt and do not respond.}. The objective of the system is to store $\nu$ independent totally ordered versions of a message of length $K$ bits. The $j$-th version of the message is denoted by $\mathbf W_j \in [2^K]$, where $j \in [\nu]$. If $i<j$, we interpret $\mathbf W_j$ as a later version with respect to $\mathbf W_i$. The $i$-th server receives an arbitrary subset of versions $\mathbf{S}(i) \subseteq [\nu]$ that denotes the \emph{state} of that server. We denote the system state by $\mathbf{S} \in \mathcal{P}([\nu])^n$, where $\mathcal{P}([\nu])$ denotes the power set of $[\nu]$. In state $\mathbf S$, we denote the set of servers that have received version $u \in [\nu]$ by $\mathcal A_{\mathbf S}(u)$.

A version that is received by at least $c_W$ servers is referred to as a \emph{complete} version. Since the system tolerates $f$ crash failures, then $c_W$ is at most $(n-f)$. A decoder that connects to an arbitrary subset of $c_R$ servers must obtain the latest complete version or a later version, where $c_R$ is at most $(n-f)$.
We provide the formal definitions next.
\begin{definition}{(Complete version).}
 In state $\mathbf S \in \mathcal P([\nu])^n$, a complete version $u \in [\nu]$ is a version that has been received by at least $c_W$ servers, that is $|\mathcal A_{\mathbf S}(u)| \geq c_W$.
\end{definition}
\noindent 
The set of complete versions in state $\mathbf S \in \mathcal P([\nu])^n$ is denoted by
\begin{align}
\mathcal C_{\mathbf S} \coloneqq \{u \in [\nu]: |\mathcal A_{\mathbf S}(u)| \geq c_W\}
\end{align}
and the latest complete version is denoted by $ L_{\mathbf S}\coloneqq\max \  \mathcal C_{\mathbf S}.$ \\
 The decoder connects to any $c_R$ servers and must decode a version $u \in [\nu]$ such that $u \geq L_{\mathbf S}$. 
We notice that among these $c_R$ servers, any complete version is present at least at $c \coloneqq c_W+c_R-n$ servers. We assume that a server is aware of the states of some other servers in the network based on the topology. Sharing the states among the servers is specified by the \emph{side information graph} that we define formally next.

\begin{definition}{(Side Information Graph).}
The side information graph $\mathcal G=(\mathcal N, \mathcal E)$ is a directed graph,  where the set of the vertices represent the servers and an edge $e_{ij} \in \mathcal E$ from vertex $i$ to vertex $j$ indicates that server $i$ is aware of the state of server $j$. 
\end{definition} 
\noindent Based on the side information graph, server $i$ obtains the states of some servers in the system. This set of servers is referred to as the neighborhood of server $i$. We next define the neighborhood of server $i$ formally.
\begin{definition}{(Neighborhood of server $i$).} The neighborhood of server $i$ is the set of servers that server $i \in \mathcal{N}$ is aware of their states which is given by 
\begin{align}
\mathcal H_i = \{j \in \mathcal N: e_{ij} \in \mathcal E \}.
\end{align}
\end{definition}
\noindent We denote the states of the servers in $\mathcal H_i$ by $\mathbf S(\mathcal H_i)$. The server stores a symbol from $[q]$ based on the versions that it receives $\mathbf W_{\mathbf S(i)}$ and the local side information $\mathbf S(\mathcal H_i)$. We next define the multi-version code with side information formally. 
\begin{definition} [Multi-version code with side information] 
	\label{Multi-version Code Definition}
	A $(\mathcal G=(\mathcal N, \mathcal E), c_W, c_R, \nu, 2^K, q)$ multi-version code with side information consists of the following 
	\begin{itemize}
		\item encoding functions
		\begin{align*}
		& \varphi_{\mathbf S(\mathcal H_i)}^{(i)} \colon {[2^K]}^{|\mathbf S(i)|} \to [q],\\ & \textit{for every}\ i \in \mathcal N \ \textit{and every}\  \mathbf S(\mathcal H_i) \subseteq \mathcal P([\nu])^{|\mathcal H_i|},
		\end{align*} 
		\item decoding functions
		\begin{align*}
		\psi_{\mathbf S}^{(\mathcal R)} \colon [q]^{c_R} &\to [2^K] \cup \lbrace \textit{NULL} \rbrace,
		\end{align*} 
		that satisfy the following  
	   \begin{align}
	  & \psi_{\mathbf S}^{(\mathcal R)} \left( \varphi_{\mathbf S(\mathcal H_{t_1})}^{(t_1)}, \cdots, \varphi_{\mathbf S(\mathcal H_{t_{c_R}})}^{(t_{c_R})} \right) =
	  \begin{cases}
	  \mathbf W_m        & \text{for some} \ m \geq L_{\mathbf S}, \\ \ & \text{if} \ \mathcal C_{\mathbf S} \neq \emptyset,  \\
	  NULL       & \text{otherwise,} 
	    \end{cases}	  
		\end{align}
		\noindent
		for every possible system state $\mathbf S \in \mathcal{P}([\nu])^n$, every $\mathbf W_{[\nu]} \in [2^K]^\nu$ and every set of servers $\mathcal R \subseteq \mathcal N$, where $\mathcal R=\lbrace t_1, t_2, \cdots, t_{c_R} \rbrace$, such that $t_1 < t_2 < \cdots < t_{c_R}$.
	\end{itemize}
\end{definition} 
\noindent The objective of the multi-version coding problem is minimizing the per-server worst-case storage cost that we define next.
\begin{definition}[Storage cost]
The storage cost of a $(\mathcal G=(\mathcal N, \mathcal E), c_W, c_R, \nu, 2^K, q)$ multi-version code is equal to $\alpha=\log q$ bits.
\end{definition} 
\noindent 
In \cite{wang2018multi}, it was shown that the storage cost for the case where the versions are independent and servers do not share their states, that is $\mathcal{H}_i=\{i\}, i \in \mathcal N$, is lower-bounded as follows 
\begin{align}
\label{old_lower_bound}
\log q \geq \frac{\nu}{c+\nu-1}K- \frac{\log \left( \nu^\nu \binom{c+\nu-1}{\nu} \right) }{(c+\nu-1)}.
\end{align}
In addition, a code construction was developed with storage cost that is given by
\begin{align}
\alpha=\max\left\lbrace \frac{\nu}{c}-\frac{(\nu-1)}{tc}, \frac{1}{t}\right\rbrace K, 
\end{align}
where
\begin{align}
t=\begin{cases}
\ceil{\frac{c-1}{\nu}}+1        & \text{if} \ c \geq (\nu-1)^2,   \\
\ceil{\frac{c}{\nu-1}}       & \text{if} \ c < (\nu-1)^2.
\end{cases}
\end{align}
We notice that if $\nu|(c-1)$, the storage cost of this code construction is given by 
\begin{align}
\alpha=\frac{\nu}{c+\nu-1}K.
\end{align}
\section{Coding With Side Information}
\label{Code Construction}
In this section, we describe achievable schemes showing that the side information can reduce the worst-case storage cost.
\begin{figure}[!htb]
	\centering
	\includegraphics[width=.95\textwidth,height=.18\textheight]{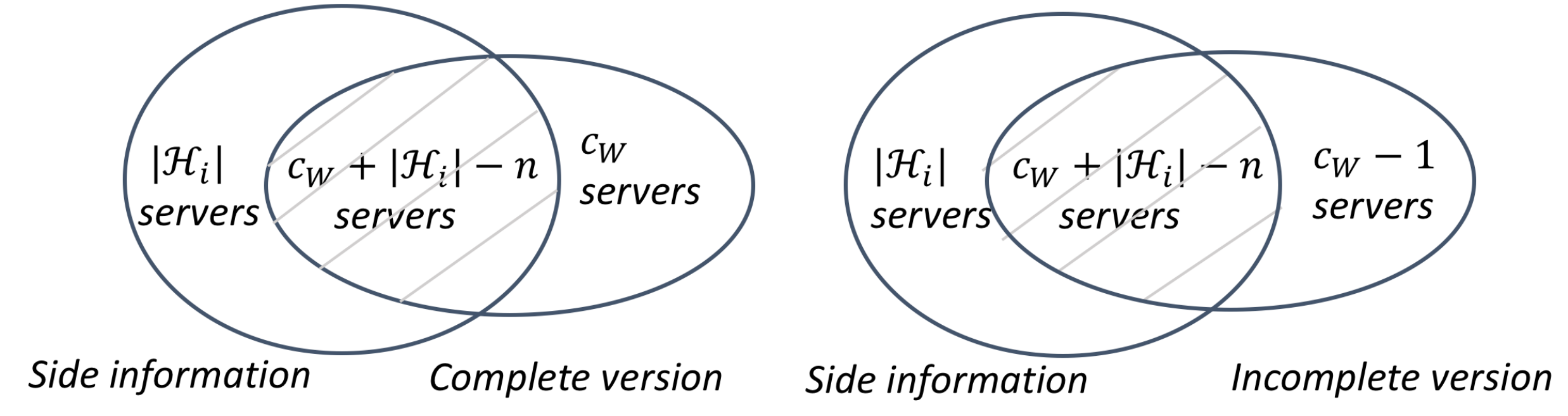}
	\caption{The intuition behind the code constructions. \label{side_information}}
\end{figure}
We begin by explaining the intuition behind the construction as shown in Fig. \ref{side_information}. We intend to provide each server a methodology to distinguish between complete and incomplete versions. Once the server can identify the complete versions, it can use its storage budget on the complete versions. Note that the greater the connectivity of the graph, the more reliably a server can ``guess'' the set of complete versions and improve the storage cost.
 
Consider a state $\mathbf S \in \mathcal P([\nu])^n$ where version $u \in [\nu]$  is complete, that is, at least $c_W$ servers have received $\mathbf W_u$. Suppose that the $i$-th server has received version $u$, i.e., $i \in \mathcal A_{\mathbf{S}}(u)$, then this server observes through its local side information at least $c_W+|\mathcal H_i|-n$ servers having $u$. However, the converse may not be true. That is, there may exist an \emph{incomplete} version $u'$ such that a server that receives $u'$ observers  $c_W+|\mathcal H_i|-n$ or more servers that receives $u'$ through its side information. In particular, this can occur when $|\mathcal{H}_i| < n,$ that is, when the graph is incomplete$^\mathrm{5}$\footnote{$^\mathrm{5}$Throughout this paper, we assume that each vertex in the side information graph has a self-edge.}. In our achievable strategy, a server that observes $c_W+|\mathcal H_i|-n$ or more neighbors having version $u$ assumes that $u$ is a complete version. For any given incomplete version $u$, the maximum number of servers that mistakenly assume that $u$ is complete is given by 
\begin{align}
\label{optimization}
\overline m_{\mathcal G}(c_W) \coloneqq  \max_{\mathbf S \in \mathcal P([\nu])^n: \mathcal A_{\mathbf S}(u) \leq (c_W-1)} \ |\{i \in \mathcal A_{\mathbf S}(u): |\mathcal A_{\mathbf S}(u) \cap \mathcal H_i| \geq c_W+|\mathcal H_i|-n \}|.
\end{align}
\noindent 
In fact, the optimization problem in (\ref{optimization}) can be performed over states which satisfy $\mathcal A_{\mathbf S}(u) = (c_W-1)$. In order to see this, consider two states $\mathbf S', \mathbf S \in \mathcal P([\nu])^n$ such that $\mathcal A_{\mathbf S'}(u) \subset \mathcal A_{\mathbf S}(u)$. In this case, we have 
 \begin{align*}
 \{i \in \mathcal A_{\mathbf S'}(u): |\mathcal A_{\mathbf S'}(u) \cap \mathcal H_i| \geq c_W+|\mathcal H_i|-n \} \subseteq \{i \in \mathcal A_{\mathbf S}(u): |\mathcal A_{\mathbf S}(u) \cap \mathcal H_i| \geq c_W+|\mathcal H_i|-n \}.
 \end{align*} 
 Therefore, the maximum corresponds to the case where $u \notin \mathcal C_{\mathbf S}$ is received by exactly $c_W-1$ servers and $\overline m_{\mathcal{G}}(c_W)$ can be simplified as follows
\begin{align}
\overline m_{\mathcal{G}}(c_W) =  \max_{\mathbf S \in \mathcal P([\nu])^n: \mathcal A_{\mathbf S}(u)=(c_W-1)} \ |\{i \in \mathcal A_{\mathbf S}(u): |\mathcal A_{\mathbf S}(u) \cap \mathcal H_i| \geq c_W+|\mathcal H_i|-n \}|.
\end{align}
We can interpret $\overline m_{\mathcal{G}}(c_W)$ in graph-theoretic terms as follows. In state $\mathbf S \in \mathcal{P}([\nu])^n$, consider a subgraph $\mathcal{G}'=(\mathcal N', \mathcal E')$ obtained by removing all vertices corresponding to the servers that do not have an incomplete version $u \notin \mathcal C_{\mathbf S}$ and the corresponding edges. By definition, the out-degree of vertex $i$ in $\mathcal G=(\mathcal N, \mathcal E)$ is given by
\begin{align*}
\mathrm{deg}_{\mathcal G}^{+}(i)=|\mathcal H_i|
\end{align*}
and the out-degree of $i$ in the induced graph $\mathcal G'=(\mathcal N', \mathcal E')$ is given by 
\begin{align*}
\mathrm{deg}_{\mathcal G'}^{+}(i)=|\mathcal A_{\mathbf S}(u) \cap \mathcal H_i|.
\end{align*}
Based on this interpretation, we can express $\overline m_{\mathcal G}(c_W)$ as follows
\begin{align}
\overline m_{\mathcal{G}}(c_W) =  \max\limits_{\mathcal{G'}=(\mathcal N', \mathcal E') \subset \mathcal G: \ |\mathcal N'|= (c_W-1) } \ |\{i \in \mathcal N': \mathrm{deg}_{\mathcal G'}^{+}(i) \geq c_W+\mathrm{deg}_{\mathcal G}^{+}(i)-n\}|.
\end{align}
\noindent 
In fact, the quantity $ \overline m_{\mathcal{G}}(c_W)$ motivates a graph functional that we term the size of \emph{the complement of the smallest $c_W$-maximally externally connected subset of a graph ($c_W$-CSMECS)}, that we define in Section \ref{Graph-theoretic Interpretation} in the context of an abstract graph $\tilde{\mathcal{G}}$. In Section \ref{Maximal externally connected subset-based code constructions}, we return to the multi-version coding with side information problem and describe code constructions whose storage cost can be derived based on the size of the $c_W$-CSMECS of the side information graph. 
\subsection{The CSMECS of a Graph}
\label{Graph-theoretic Interpretation}
In this subsection, we develop the concept of maximally externally connected subset of a graph $\tilde {\mathcal G}=(\tilde{\mathcal V}, \tilde{\mathcal E}).$ We use the tilde notation here to distinguish an arbitrary graph used in this subsection from the side information graph $\mathcal{G}$ that arises in the multi-version coding problem. 

Consider a directed graph $\tilde {\mathcal G}=(\tilde{\mathcal V},\tilde{\mathcal E})$. A vertex $u \in \tilde{\mathcal V}$ is an in-neighbor of a vertex $v \in \tilde{\mathcal V}$ if $(u, v) \in \tilde{\mathcal E}$ and an out-neighbor of $v$ if $(v, u) \in \tilde{\mathcal E}$. We use $\mathcal N^+_{\tilde {\mathcal G}}(v)$ to denote the set of out-neighbors and $\mathcal N^-_{\tilde {\mathcal G}}(v)$ to denote the set of in-neighbors of $v \in \tilde{\mathcal V}$. An induced subgraph $\tilde {\mathcal G}[\tilde{\mathcal V'}] \subseteq \tilde {\mathcal G}$ is a subgraph of $\tilde {\mathcal G}$ formed by a subset of the vertices $\tilde{\mathcal V'} \subseteq \tilde{\mathcal V}$ and all of the edges connecting pairs of vertices in $\tilde{\mathcal V}'$. The subgraph induced by the in-neighborhood  of a vertex $v \in \tilde{\mathcal V}$ is called the in-neighborhood graph of $v$ and is denoted by $\tilde {\mathcal G}[\mathcal N^-_{\tilde {\mathcal G}}(v)]$. The subgraph induced by the out-neighborhood  of a vertex $v \in \tilde{\mathcal V}$ is called the out-neighborhood graph of $v$ and is denoted by $\tilde {\mathcal G}[\mathcal N^{+}_{\tilde {\mathcal G}}(v)]$.

Consider a graph $\tilde {\mathcal G}=(\tilde{\mathcal V}, \tilde{\mathcal E})$ and any induced subgraph $\tilde{\mathcal G}[\tilde{\mathcal U}]$, where $\tilde{\mathcal U} \subseteq \tilde{\mathcal{V}}$. The out-degree of any vertex $u \in \tilde{\mathcal U}$ is lower-bounded as follows 
\begin{align}
\label{out-degree lower bound}
\mathrm{deg}^{+}_{\tilde {\mathcal G}[\tilde{\mathcal U}]}(u)& \coloneqq | \mathcal N^{+}_{\tilde {\mathcal G}}(u) \cap \tilde{\mathcal U}|  \notag \\ 
&= |\mathcal N^{+}_{\tilde {\mathcal G}}(u)|+|\tilde{\mathcal U}| -|\mathcal N^{+}_{\tilde {\mathcal G}}(u) \cup \tilde{\mathcal U}| 
\notag \\ & \geq \mathrm{deg}^{+}_{\tilde {\mathcal G}}(u)+|\tilde{\mathcal U}|-|\tilde{\mathcal V}|.
\end{align}

Given a subgraph $ \tilde {\mathcal G'}=(\tilde{\mathcal V'}, \tilde{\mathcal E'}) \subset \tilde{\mathcal G}$, where $|\tilde{\mathcal V'}|=(s-1)$, an $s$-maximally externally connected subset ($s$-MECS) is a subset of $\tilde {\mathcal V'}$ such that each vertex is connected to more than $|\tilde{\mathcal V}|-s$ vertices outside $\tilde{\mathcal V'},$ that is in $\tilde{\mathcal{V}} - \tilde{\mathcal V'}$. That is, an $s$-MECS is the set of all vertices such that each connected to every vertex outside $\tilde{\mathcal V'}$. The \emph{$s$-SMECS of a graph $\tilde{\mathcal G}$} is the \emph{smallest} $s$-MECS, where the minimization is over all possible subgraphs $\tilde {\mathcal G'}=(\tilde{\mathcal V'}, \tilde{\mathcal E'}) \subset \tilde{\mathcal G}$, where $|\tilde{\mathcal V'}|=(s-1)$. \\ We now define \emph{the size of the $s$-SMECS of a graph} formally.
\begin{definition}[Size of the Smallest $s$-Maximally Externally Connected Subset ($s$-SMECS) of a Graph]
In a graph $\tilde {\mathcal G}=(\tilde{\mathcal V}, \tilde{\mathcal E})$, for every integer $1 \leq s \leq |\tilde{\mathcal{V}}|,$ the size of the smallest $s$-maximally externally connected subset  ($s$-SMECS) of $\mathcal G$, $m_{\tilde {\mathcal G}}(s)$, is given by
\begin{align}
\label{maximum graph}
m_{\tilde {\mathcal G}}( s) =   \min \limits_{\tilde {\mathcal G}' = (\tilde{\mathcal{V}}',\tilde{\mathcal{E}}') \subset \tilde {\mathcal G}: \ |\tilde{\mathcal V'}|= (s-1) } \ |\{u \in \tilde{\mathcal V'}: \mathrm{deg}^{+}_{\tilde {\mathcal G}}(u)-\mathrm{deg}^{+}_{\tilde {\mathcal G}'}(u) >  |\tilde{\mathcal V}|-s\}|.
\end{align}	
\end{definition}
It is instructive to note that the size of the $s$-SMECS of a graph does not change if the minimization in its definition is performed over all sub-graphs $\tilde{\mathcal{G}'}=(\tilde{\mathcal{V}}',\tilde{\mathcal{E}}')$, where $|\tilde{\mathcal{V}'}| \leq s-1$. \\
\noindent We denote the size of the complement of the $s$-SMECS set by $\overline m_{\tilde {\mathcal G}}( s)$ and refer to it as the $s$-CSMECS of a graph. The $s$-CSMECS is an important graph-theoretic quantity and dictates the storage costs of our code constructions. Note that the $s$-CSMECS of the graph can be equivalently written as 
\begin{align}
\overline m_{\tilde {\mathcal G}}( s) &=  \max \limits_{\tilde {\mathcal G}' = (\tilde{\mathcal{V}}',\tilde{\mathcal{E}}') \subset \tilde {\mathcal G}: \ |\tilde{\mathcal V'}|= (s-1) } \ |\{u \in \tilde{\mathcal V'}:    \mathrm{deg}^{+}_{\tilde {\mathcal G}}(u)-\mathrm{deg}^{+}_{\tilde {\mathcal G}'}(u) \leq |\tilde{\mathcal V}|-s \}|.\\
&= \max \limits_{\tilde {\mathcal G}' = (\tilde{\mathcal{V}}',\tilde{\mathcal{E}}') \subset \tilde {\mathcal G}: \ |\tilde{\mathcal V'}|= (s-1) } \ |\{u \in \tilde{\mathcal V'}:   \mathrm{deg}^{+}_{\tilde {\mathcal G}'}(u) \geq  \mathrm{deg}^{+}_{\tilde {\mathcal G}}(u)-|\tilde{\mathcal V}|+s \}|.\\
&= (s-1)-m_{\tilde {\mathcal G}}( s).
\end{align}

Before we describe our constructions and associated storage costs, we give examples to illustrate the calculation of the $s$-CSMECS of the graph $\overline{m}_{\tilde {\mathcal G}}( s)$ for various graphs and provide some bounds. In all of our examples and throughout this paper, we assume that the vertices have self edges and we do not show them. 
\begin{example}
\label{Example_general_graph}	
Consider the graph $\tilde{\mathcal G}=(\tilde{\mathcal V}, \tilde{\mathcal E})$ shown in Fig. \ref{Example_general_graph_figure}, where $\tilde{\mathcal V}=\{0, 1, 2, 3, 4\}$. Consider the case where $s=5$ and consider a subgraph $\tilde{\mathcal G}'=(\tilde{\mathcal V'}, \tilde{\mathcal E'})$, where $\tilde{\mathcal V'}=\tilde{\mathcal V} \setminus \{4\}$.
	\begin{figure}[h]
	\centering
	\includegraphics[width=.5\textwidth,height=.22\textheight]{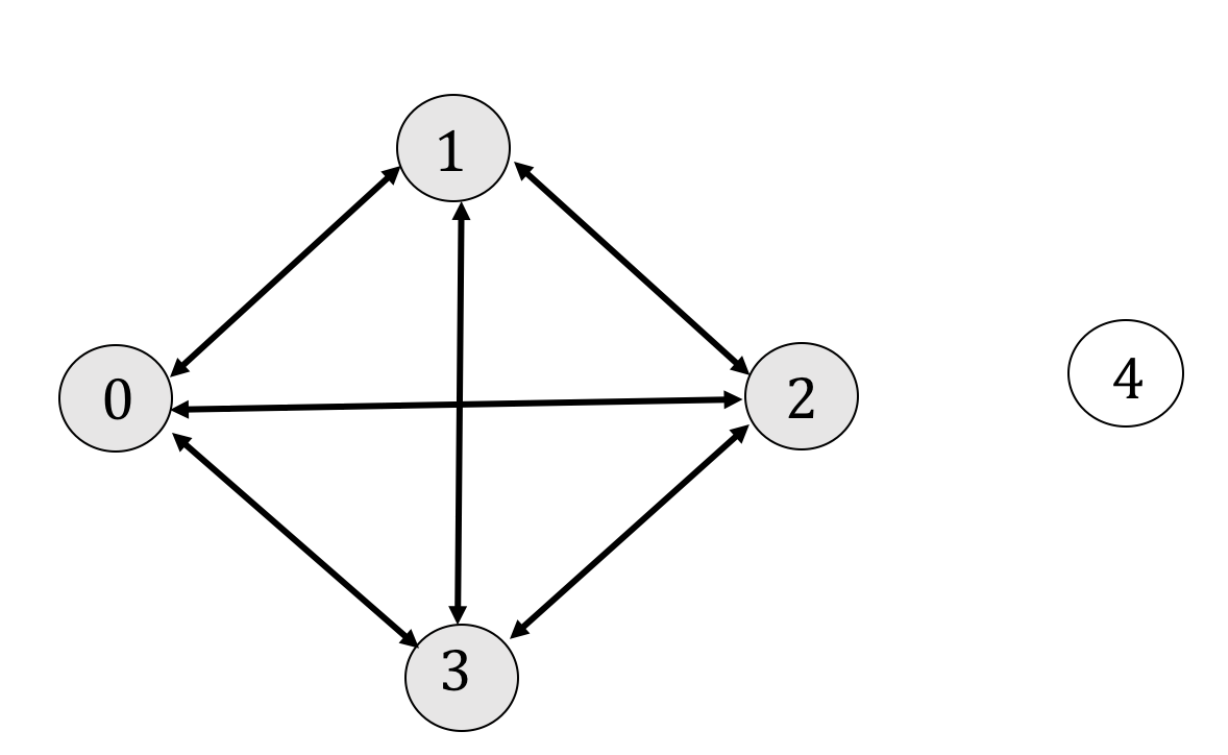}
	\caption{A graph $\tilde{\mathcal G}=(\tilde{\mathcal V}, \tilde{\mathcal E})$, where  $|\tilde{\mathcal V}|=\{0, 1, 2, 3,4\}$. For the subgraph $\tilde{\mathcal G}'=(\tilde{\mathcal V'}, \tilde{\mathcal E'})$, where $\tilde{\mathcal V'}=\tilde{\mathcal V} \setminus \{4\}$, the vertices $0, 1, 2, 3$ still have an out-degree of $4$ and hence $m_{\tilde {\mathcal G}}(s)=0$. \label{Example_general_graph_figure}}
\end{figure}

In this case, we have
\begin{align*}
  m_{\tilde {\mathcal G}}(5)=|\{u \in \tilde{\mathcal V'}: \mathrm{deg}^{+}_{\tilde {\mathcal G}}(u) - \mathrm{deg}^{+}_{\tilde {\mathcal G}'}(u) > 0\}|=0,
\end{align*}
and hence the $5$-CSMECS is $\overline{{m}}_{\tilde{\mathcal{G}}}(5)=4$. 
\end{example}

In general, determining $\overline{m}_{\tilde{\mathcal G}}(s)$ is a discrete optimization that is computationally intractable for large graphs. We derive an upper bound on $ \overline{m}_{\tilde{\mathcal G}}(s)$ for regular directed graphs in Lemma \ref{lower-bound lemma}. We recall that in $k$-regular directed graph $\tilde {\mathcal G}$ every vertex has in-degree as well as out-degree equal to $k$. For such a graph, we have
\begin{align}
 \overline{m}_{\tilde{\mathcal G}}(s) =  \max \limits_{\tilde {\mathcal G}' = (\tilde{\mathcal{V}}',\tilde{\mathcal{E}}') \subset \tilde {\mathcal G}: \ |\tilde{\mathcal V'}|= (s-1)} \ |\{u \in \tilde{\mathcal V'}:   \mathrm{deg}^{+}_{\tilde {\mathcal G}'}(u) > |\tilde{\mathcal V}|+k-s\}|.
\end{align}
We first consider the following useful lemma. 

\begin{lemma}
	\label{nescessary and sufficient condition}
	Consider any subgraph $\tilde{\mathcal{G}'}=(\tilde{\mathcal{V}'},\tilde{\mathcal{E}'})$ of a $k$-regular graph $\tilde{\mathcal{G}}=(\tilde{\mathcal{V}}, \tilde{\mathcal{E}})$, where $|\tilde{\mathcal{V}'}|=s-1$. For any vertex a vertex $u \in \tilde{\mathcal V'}$, $\mathrm{deg}^{+}_{\tilde {\mathcal G}'}(u) \geq k+s-|\tilde{\mathcal V}|$ if and only if there exists at least one vertex $i \in \tilde{\mathcal{V}}$ such that $i \notin  \mathcal N^{+}_{\tilde {\mathcal G}}(u) \cup \tilde{\mathcal V'}$.
	
\end{lemma}

\begin{proof}
	Consider a vertex $u \in \tilde{\mathcal V'}$. Suppose there exists at least one vertex $i \notin  \mathcal N^{+}_{\tilde {\mathcal G}}(u) \cup \tilde{\mathcal V'}$, then we have 
	\begin{align*}
	\mathrm{deg}^{+}_{\tilde {\mathcal G}'}(u)& = | \mathcal N^{+}_{\tilde {\mathcal G}}(u) \cap \tilde{\mathcal V'}| 
	\notag \\ 
	&=| \mathcal N^{+}_{\tilde {\mathcal G}}(u) |+|\tilde{\mathcal V'}|-|\mathcal N^{+}_{\tilde {\mathcal G}}(u) \cup \tilde{\mathcal V'}|
	\notag \\
	& \geq k+(s-1)-(|\tilde{\mathcal V}|-1) \notag \\
	&=k+s-|\tilde{\mathcal V}|.
	\end{align*}
	Conversely, suppose that $\mathcal N^{+}_{\tilde {\mathcal G}}(u) \cup \tilde{\mathcal V'}= \tilde{\mathcal V}$. In this case, we have
		\begin{align*}
	\mathrm{deg}^{+}_{\tilde {\mathcal G}'}(u)& = | \mathcal N^{+}_{\tilde {\mathcal G}}(u) \cap \tilde{\mathcal V'}| 
	\notag \\ 
	&=| \mathcal N^{+}_{\tilde {\mathcal G}}(u) |+|\tilde{\mathcal V'}|-|\mathcal N^{+}_{\tilde {\mathcal G}}(u) \cup \tilde{\mathcal V'}|
	\notag \\
	&=k+(s-1)-|\tilde{\mathcal V}|.
	\end{align*}
	
\end{proof}
\noindent We now provide a lower bound on  $\overline{m}_{\tilde{\mathcal G}}(s)$ for $k$-regular graphs in Lemma \ref{lower-bound lemma}.

\begin{lemma}
	\label{lower-bound lemma}
	For any $k$-regular graph $\tilde {\mathcal G}$, we have	
	\begin{align}
	 \overline{m}_{\tilde{\mathcal G}}(s) \leq \min \big( (|\tilde{\mathcal V}|-s+1)(|\tilde{\mathcal V}|-k), s-1 \big) .
	\end{align}
\end{lemma}
\begin{proof}
We can upper-bound $\overline m_{\tilde{\mathcal G}}(s)$ as follows
	\begin{align*}
\overline m_{\tilde{\mathcal G}}(s) &=  \max\limits_{\tilde {\mathcal G}' \subset \tilde {\mathcal G}: \ |\tilde{\mathcal V'}|= (s-1) } \ |\{u \in \tilde{\mathcal V'}: \mathrm{deg}^{+}_{\tilde {\mathcal G}'}(u) \geq k+s-|\tilde{\mathcal V}|\}| \notag\\
	&\stackrel{(a)} =  \max\limits_{\tilde {\mathcal G}' \subset \tilde {\mathcal G}: \ |\tilde{\mathcal V'}|= (s-1) } \big| \bigcup_{i \in \tilde{\mathcal V} \setminus \tilde{\mathcal V'}} 
	\{u \in \tilde{\mathcal V'}: i \notin \mathcal N^{+}_{\tilde {\mathcal G}}(u)\} \big|  
	\notag \\ & \stackrel{(b)} \leq  \max\limits_{\tilde {\mathcal G}' \subset \tilde {\mathcal G}: \ |\tilde{\mathcal V'}|= (s-1) } \sum_{i \in \mathcal V \setminus \tilde{\mathcal V'}} |	\{u \in \tilde{\mathcal V}: i \notin \mathcal N^{+}_{\tilde {\mathcal G}}(u)\}| 
	\notag \\ 
	& \stackrel{(c)} \leq  \max\limits_{\tilde {\mathcal G}'\subset \tilde {\mathcal G}': \ |\tilde{\mathcal V'}|= (s-1) } \sum_{i \in \tilde{\mathcal V} \setminus \tilde{\mathcal V'}} (|\tilde{\mathcal V}|-k) \\
	& = (|\tilde{\mathcal V}|-s+1)(|\tilde{\mathcal V}|-k),
	\end{align*} 
	where $(a)$ follows from Lemma \ref{nescessary and sufficient condition}, $(b)$ follows by the union bound and $(c)$ follows since the graph is $k$-regular.
	
\end{proof}
Now, we consider the two extreme cases of the graph $\tilde {\mathcal G}=(\tilde{\mathcal V}, \tilde{\mathcal E})$. The first case corresponds to a $|\tilde{\mathcal V}|$-regular graph $\tilde {\mathcal G}$, that is, the graph is a clique. In this case, we have  
\begin{align*}
\overline{m}_{\tilde {\mathcal G}}(s) &= (s-1)-\min \limits_{\tilde {\mathcal G}'=(\tilde{\mathcal V'}, \tilde{\mathcal E'}) \subset \tilde {\mathcal G}': \ |\tilde{\mathcal V'}|= (s-1) }\  |\{u \in \tilde{\mathcal V'}: \mathrm{deg}_{\tilde {\mathcal G}'}^{+}(u) < s\}| =0
\end{align*}

In the second case, there are no edges between the different vertices in $\tilde {\mathcal G}$ and the graph has only self edges. In this case, we have  
\begin{align*}
\overline{m}_{\tilde {\mathcal G}}(s) &= (s-1)-\min \limits_{\tilde {\mathcal G}' = (\tilde{\mathcal{V}}',\tilde{\mathcal{E}}') \subset \tilde {\mathcal G}: \ |\tilde{\mathcal V'}|= (s-1)} \ |\{u \in \tilde{\mathcal V'}: \mathrm{deg}^{+}_{\tilde {\mathcal G}'}(u) < 1+s-|\tilde{\mathcal V}|\}|=s-1.
\end{align*}
In Example \ref{Upper bound example1}, Example \ref{Upper bound example2}, we provide two cases of regular graphs that match the upper bound obtained in Lemma \ref{lower-bound lemma}. 
\begin{example}
\label{Upper bound example1}	
Consider a $k$-regular graph $\tilde{\mathcal G}=(\tilde{\mathcal V}, \tilde{\mathcal E})$, where $k=5$ and $\tilde{\mathcal V}=\{0, 1, 2, \cdots, 11\}$ as shown in Fig. \ref{upper_bound_example1_figure1}. Consider that case where $s=|\tilde{\mathcal V}|=12$ and consider a subgraph $\tilde{\mathcal G}'=(\tilde{\mathcal V'}, \tilde{\mathcal E'})$, where $\tilde{\mathcal V'}=\tilde{\mathcal V} \setminus \{0\}$ as shown in Fig. \ref{upper_bound_example1_figure2}. In this case, we have
\begin{align*}
|\{u \in \tilde{\mathcal V'}: \mathrm{deg}^{+}_{\tilde {\mathcal G}'}(u) < k+s-|\tilde{\mathcal V}|\}| &=|\{u \in \tilde{\mathcal V'}: \mathrm{deg}^{+}_{\tilde {\mathcal G}'}(u) < 5\}| \notag \\ 
&=|\{1, 2, 9, 10 \}| =4,
\end{align*}
and hence $\overline{m}_{\tilde{\mathcal{G}}}(12)=7,$  which matches the upper bound of Lemma \ref{lower-bound lemma}.
	\begin{figure}[h]
		\centering
		\includegraphics[width=.42\textwidth,height=.25\textheight]{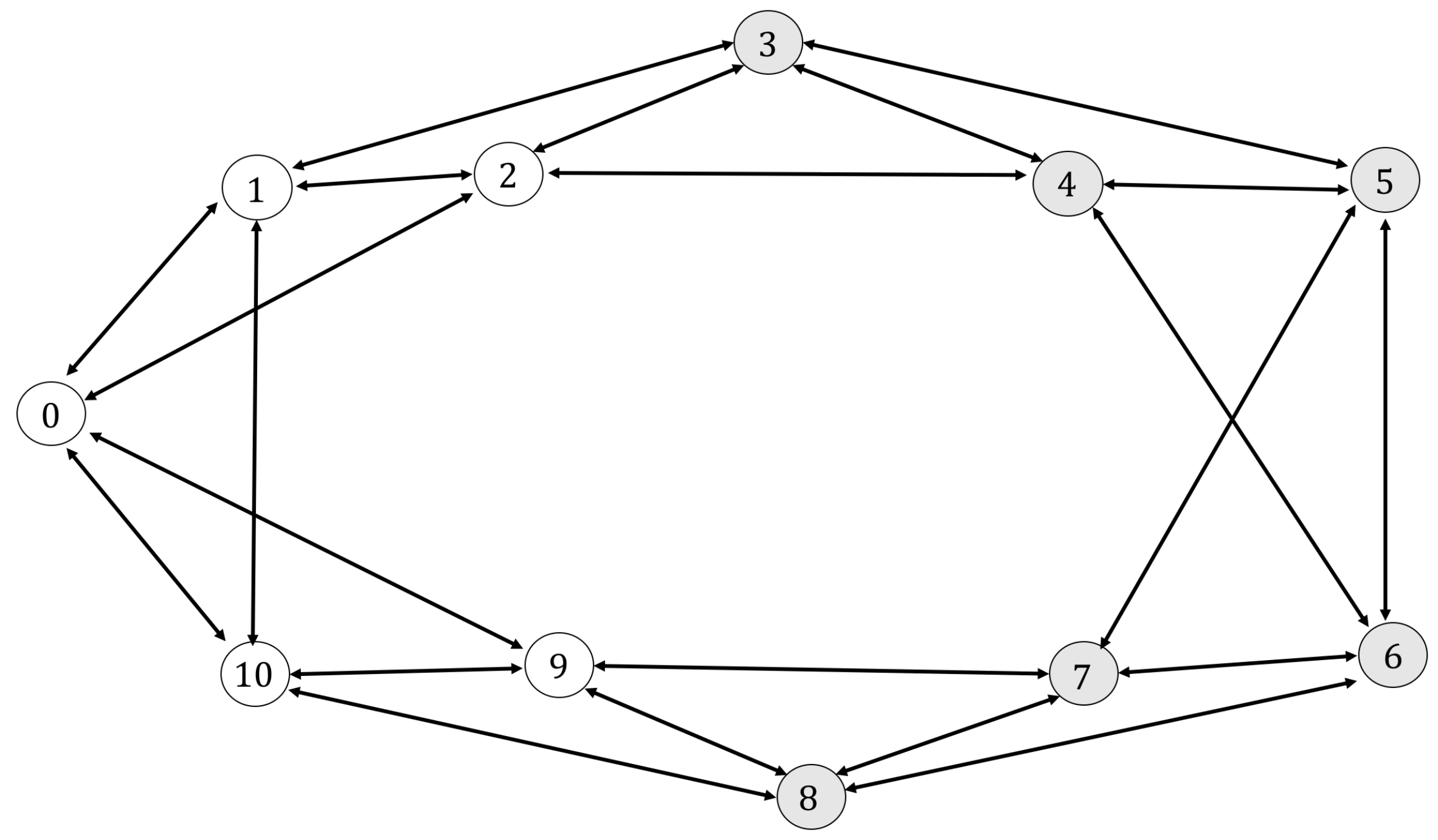}
		\caption{A $k$-regular graph $\tilde{\mathcal G}=(\tilde{\mathcal V}, \tilde{\mathcal E})$ with $k=5$ and $|\tilde{\mathcal V}|=11$. \label{upper_bound_example1_figure1}}
	\end{figure}
	\begin{figure}[h]
	\centering
	\includegraphics[width=.42\textwidth,height=.25\textheight]{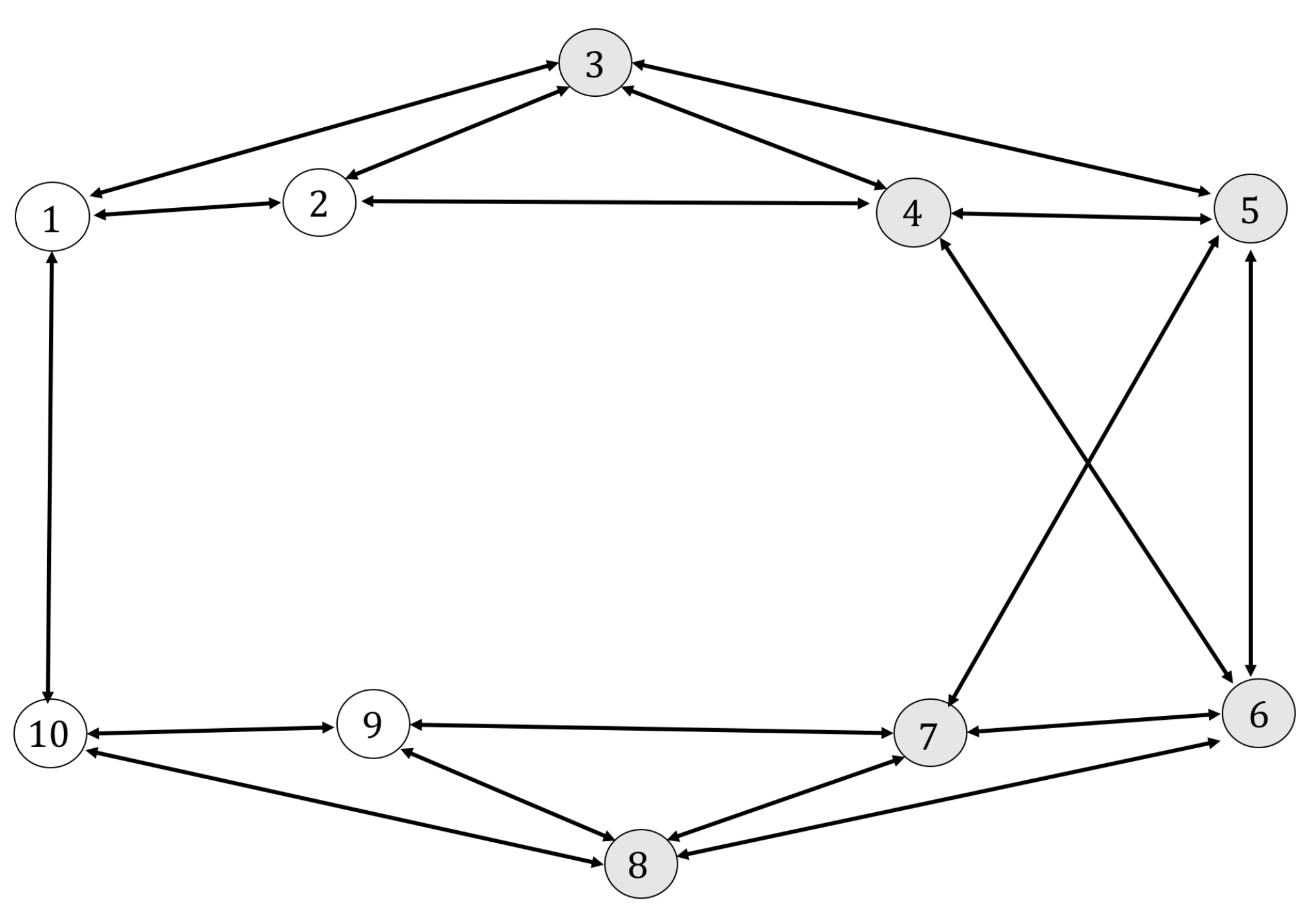}
	\caption{A subgraph $\tilde{\mathcal G}'=(\tilde{\mathcal V'}, \tilde{\mathcal E'})$, where $\tilde{\mathcal V'}=\tilde{\mathcal V} \setminus \{0\}$. Vertices $1, 2, 9, 10$ now have an out-degree $4$ after the removal of vertex $0$. \label{upper_bound_example1_figure2}}
\end{figure}
\end{example}

\begin{example}
	\label{Upper bound example2}	
	Consider a $k$-regular graph $\tilde{\mathcal G}=(\tilde{\mathcal V}, \tilde{\mathcal E})$, where $k=9$ and $\tilde{\mathcal V}=\{0, 1, \cdots, 10\}$. The out-neighborhood of $v \in \tilde{\mathcal V}$ is given by
	\begin{align*}
	\mathcal N^{+}_{\tilde{\mathcal G}}(v)=\{v-(k-1)/2, \cdots, v-2, v-1,v, v+1, v+2, \cdots, v+(k-1)/2\},
	\end{align*}
	where the addition is modulo $|\tilde{\mathcal V}|$. Consider the case where $s=9$ and consider a subgraph $\tilde{\mathcal G}'=(\tilde{\mathcal V'}, \tilde{\mathcal E'})$, where $\tilde{\mathcal V'}=\tilde{\mathcal V} \setminus \{0, 3, 7\}$ as shown in Fig. \ref{upper_bound_example2_figure2}. In this case, we have
	\begin{align*}
	m_{\tilde{\mathcal{G}}}(9) = |\{u \in \tilde{\mathcal V'}: \mathrm{deg}^{+}_{\tilde {\mathcal G}'}(u) < k+s-|\tilde{\mathcal V}|\}|&=|\{u \in \tilde{\mathcal V'}: \mathrm{deg}^{+}_{\tilde {\mathcal G}'}(u) < 7\}| \notag \\
	&=|\{4, 10\}|=2,
	\end{align*} 
which implies that $\overline{m}_{\tilde{\mathcal{G}}}(9)=6,$ which  matches the upper bound of Lemma \ref{lower-bound lemma}.	
	\begin{figure}[h]
		\centering
		\includegraphics[width=.42\textwidth,height=.24\textheight]{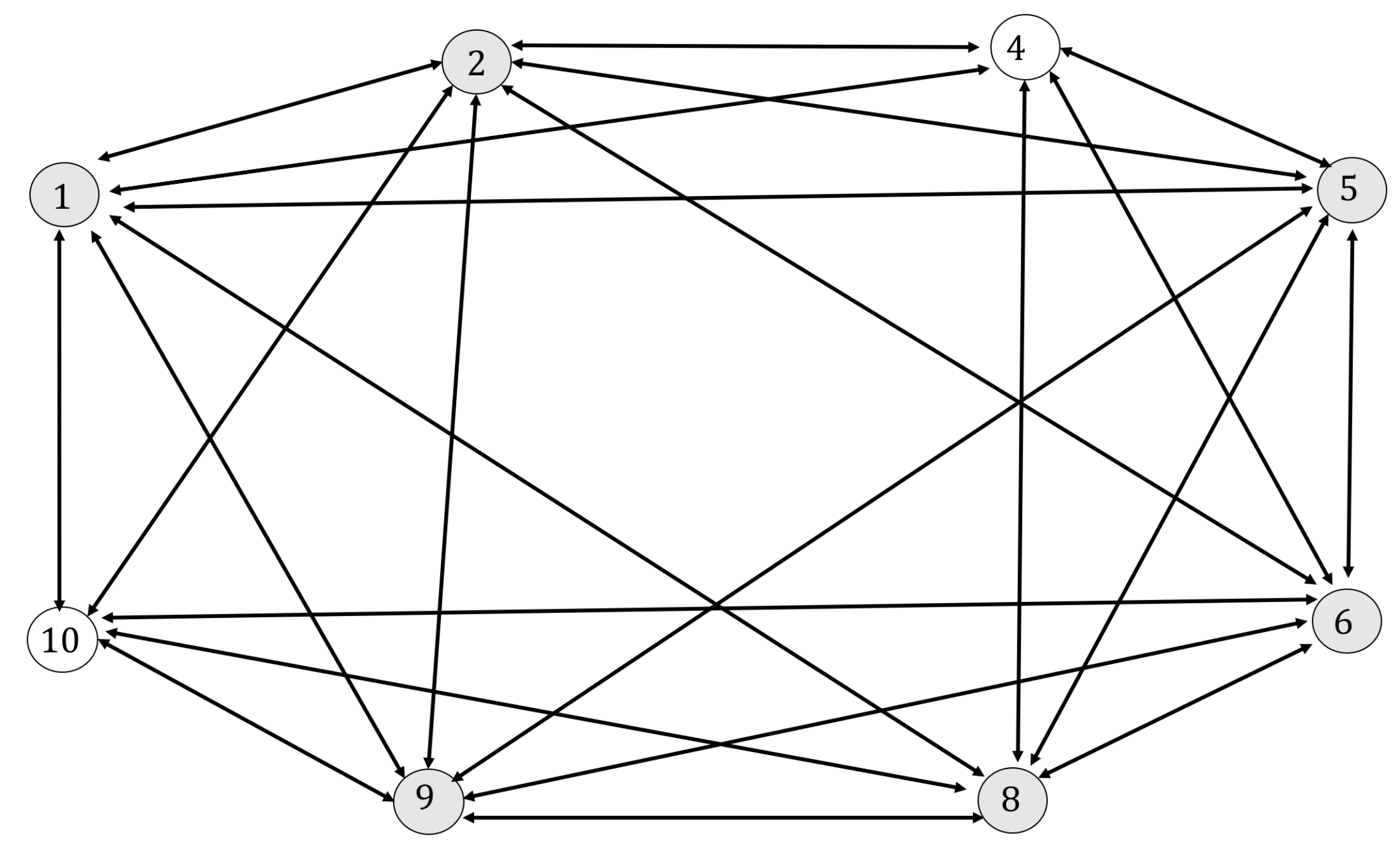}
		\caption{A subgraph $\tilde {\mathcal G}'=(\tilde{\mathcal V'}, \tilde{\mathcal E'})$, where $\tilde{\mathcal V'}=\tilde{\mathcal V} \setminus \{0, 3, 7\}$. Vertices $4, 10$ have out-degree of $6$. \label{upper_bound_example2_figure2}}
	\end{figure}
	
\end{example}

\subsection{Code Constructions }
\label{Maximal externally connected subset-based code constructions}
 We begin with some notations. Consider the $i$-th server, where $i \in \mathcal N$, and a system state $\mathbf S \in \mathcal P([\nu])^n$. The $i$-th server stores $\alpha_{i, u}^{\mathbf S(\mathcal{H}_i)}$ bits of version $u$ in this state, where $u\in [\nu]$. The worst-case storage cost is then given by 
\begin{align}
\alpha= \max\limits_{i, \mathbf S}
\sum\limits_{u=1}^{\nu} \alpha_{i, u}^{\mathbf S(\mathcal H_i)}.
\end{align} 
\noindent In state $\mathbf S \in \mathcal P([\nu])^n$, we denote the latest version that server $i$ receives that is at least received by $c_W+|\mathcal H_i|-n$ servers in the neighborhood of server $i$ by
 \begin{align}
 L_{\mathbf S(i)} = \max \ \{u \in \mathbf S(i): |\mathcal A_{\mathbf S}(u) \cap \mathcal H_i| \geq c_W+|\mathcal H_i|-n \}.
 \end{align} 
We now provide our first construction. In this construction, when the $i$-th server observes at least $c_W+|\mathcal H_i|-n$ servers having a $\mathbf W_2$, it stores $K/c$ of $\mathbf W_2$. Since observing $c_W+|\mathcal H_i|-n$ servers having $\mathbf W_2$ does not imply that $\mathbf W_2$ is a complete version, the $i$-th server allocates the remaining storage budget of $(\alpha-K/c)$ to $\mathbf W_1$ as it may be the latest complete version. We provide the construction formally next. 
\begin{construction}
\label{first construction}
We construct a code as follows for $\nu=2$
\begin{align}
\alpha= \frac{c+ \overline m_{\mathcal G}(c_W) }{c^2} K
\end{align}
and 
\begin{align}
 &\alpha_{i, 2}^{\mathbf S(\mathcal H_i)}= 
\begin{cases}
K/c &\text{$L_{\mathbf S(i)}=2,$}\\
0 &\text{otherwise,}
\end{cases} \\
& \alpha_{i, 1}^{\mathbf S(\mathcal H_i)}=\alpha-\alpha_{i, 2}^{\mathbf S(\mathcal H_i)},
\end{align}
where $c_W+|\mathcal H_i|\geq n, \forall i \in \mathcal N$.
\end{construction}
\begin{theorem}
\label{first construction theorem}	
Construction \ref{first construction} is a $(\mathcal G=(\mathcal N, \mathcal E), c_W, c_R, \nu=2, 2^K, q)$ multi-version code with side information with a worst-case storage cost of 
\begin{align}
 \frac{c+\overline m_{\mathcal G}(c_W)}{c^2} K,
\end{align}
where $c_W+|\mathcal H_i|\geq n, \forall i \in \mathcal N$.
\end{theorem}
\begin{proof}
Consider any state $\mathbf S \in \mathcal P([\nu])^n$. We show that the latest complete version in this state, version $L_{\mathbf S}$, is decodable. Specifically, we consider the following cases.
\begin{itemize}
	\item \textbf{Case $1$} ($L_{\mathbf S}=1$). Since $L_{\mathbf S}=1$, then version $2$ is incomplete. In this case, at most $\overline m_{\mathcal G} (c_W)$ servers will allocate a storage budget to version $2$. According to the construction, each of these servers will store $K/c$ of version $2$ and $(\alpha-K/c)$ of version $1$. Therefore, the storage allocation of version $1$ is at least
	\begin{align*}
	\overline m_{\mathcal G}(c_W)(\alpha-K/c) + (c-\overline m_{\mathcal G}(c_W)) \alpha=K. 
	\end{align*}
Therefore, version $1$ is decodable in this state. 	
	\item \textbf{Case $2$} ($L_{\mathbf S}=2$). Since version $2$ is complete, then at least $c_W$ servers have it. The decoder connects to any $c_R$ servers. Among these $c_R$ servers, at least $c$ servers have received this version. Denote this set of servers by $T=\{t_1, t_2, \cdots, t_c\} \subseteq \mathcal N$. Server $t_i \in T, i \in [c],$ observes at least $c_W+|\mathcal H_{t_i}|-n$ servers that have received version $2$ and hence it stores $K/c$ of version $2$. Since each of the $c$ servers stores $K/c$ of versions $2$, then versions $2$ is decodable in this state. 
\end{itemize}
\end{proof}
\begin{remark}
For network topologies where $\overline m_{\mathcal G}(c_W)<\frac{c(c-1)}{c+1}$, the storage cost of Theorem \ref{first construction theorem} is strictly less than $\frac{2}{c+1}K$ and hence the side information is useful in those cases.  
\end{remark}

\noindent We next provide our second construction for any number of versions $\nu$. 
\begin{construction}
	\label{second construction}
	We construct a code as follows 
	\begin{align}
	\alpha=\frac{1}{c-(\nu-1) \overline m_{\mathcal G}(c_W)} K,
	\end{align}
	\begin{align}
	&\alpha_{i, u}^{\mathbf S(\mathcal H_i)}= 
	\begin{cases}
	\alpha  &\text{if} \ L_{\mathbf S(i)}=u,\\ 
	 0 &\text{otherwise,}
	\end{cases} 
	\end{align}	
where $i \in \mathcal N$ and $c > (\nu-1) \overline m_{\mathcal G}(c_W)$.
\end{construction}
\begin{theorem}
	\label{second construction theorem}	
	Construction \ref{second construction} is a $(\mathcal G=(\mathcal N, \mathcal E), c_W, c_R, \nu, 2^K, q)$ multi-version code with side information with a worst-case storage cost of 
	\begin{align}
	\frac{1}{c-(\nu-1) \overline m_{\mathcal G}(c_W)} K,
	\end{align}
where $c >  (\nu-1) \overline m_{\mathcal G}(c_W)$.	
\end{theorem}
\begin{proof}
Consider any state $\mathbf S \in \mathcal P([\nu])^n$. We show that the latest complete version in this state, version $L_{\mathbf S}$, is decodable. Specifically, we have the following cases.  
\begin{itemize}
 \item Consider any other version $u$ such that $u<L_{\mathbf S}$. Suppose that the decoder connects to any set of $c_R$ servers. Among these $c_R$ servers there are at least $c$ servers, denoted by $T=\{t_1, t_2, \cdots, t_c \}$, that have version $L_{\mathbf S}$. For any server $t_i \in T, i \in [c],   L_{\mathbf S} \leq L_{\mathbf S(t_i)}$. Therefore, for any server $t_i \in T, i \in [c]$, we have $u < L_{\mathbf S(t_i)}$ and hence none of these $c$ servers store $u$ according to the construction.
\item Consider any other version $u > L_{\mathbf S}$. Since version $u$ is incomplete, we have 
\begin{align*}
\ |\{i \in \mathcal A_{\mathbf S}(u): |\mathcal A_{\mathbf S}(u) \cap \mathcal H_i| \geq c_W+|\mathcal H_i|-n \}| \leq \overline m_{\mathcal G}( c_W). 
\end{align*}
A server $i$ of those servers stores $\alpha$ of version $u$ if $u=L_{\mathbf S(i)}$. Since there are at most $(\nu-1)$ versions that are not equal to $L_{\mathbf S}$, at least $c-(\nu-1) \overline m_{\mathcal G}(c_W)$ servers will store $\frac{1}{c-(\nu-1) \overline m_{\mathcal G}(c_W)} K$ of version $L_{\mathbf S}$. Therefore, the storage allocation of version $L_{\mathbf S}$ is at least 
\begin{align*}
(c-(\nu-1) \overline m_{\mathcal G}(c_W)) \frac{1}{c-(\nu-1) \overline m_{\mathcal G}(c_W)} K=K.
\end{align*}
\end{itemize}
Therefore, we conclude that version $L_{\mathbf S}$ is decodable in this state. 
\end{proof}
\begin{remark}
Construction $\ref{first construction}$ has a strictly better storage cost as compared with Construction \ref{second construction} for the case where $\nu=2$ for $\overline m_{\mathcal G}(c_W) >0$.
\end{remark}
\begin{remark}
	For network topologies where $\overline m_{\mathcal G}(c_W)<\frac{c-1}{\nu}$, the storage cost of Theorem \ref{second construction theorem} is strictly less than $\frac{\nu}{c+\nu-1}K$ and hence the side information is useful in those cases.  
\end{remark}
We compare between the case where there is no side information, the case where there is a partial side information and the complete side information case in Table \ref{table: comparison} in terms of the storage cost.
\begin{table*}[h]
\centering	
\begin{tabular}{ | c | c |}
\hline
Case & Storage cost 
\\  [1ex] \hline   
No side information & $\alpha \geq (\frac{\nu}{c} - \frac{\nu(\nu-1)}{c^2}+o(\frac{1}{c^2})) K$
\\  [2ex] \hline    
Partial side information & $\alpha=\left( \frac{1}{c}+\frac{(\nu-1) \overline m_{\mathcal G}( c_W)}{c^2}+o\left(\frac{\overline m_{\mathcal G}(c_W)}{c^2} \right) \right)K$ 
\\ [2 ex] \hline
Complete side information & $\alpha=\frac{1}{c}K$
\\ \hline 
\end{tabular}
\caption{The storage costs and the side information communication costs for the various side information regimes.}.\label{table: comparison}
\end{table*}

\section{Lower Bounds on the storage costs}
\label{Impossibility Results}
In this section, we provide lower bounds on the storage cost. In Theorem \ref{theorem: stronger general impossibility}, we provide a lower bound on the storage cost for any general topology that satisfies certain condition. Later on, in Theorem \ref{multi-hop Theorem: stronger impossibility}, we provide a lower bound on the storage cost for a  symmetric multi-hop topology where the servers are distributed in a ring such that each server is aware of the states of its $h$-hop neighbors. 

\subsection{Lower bound for general topology}
We begin by studying a general side information topology such that there are $c-a$ servers that are not aware of the states of other $a$ servers, where $c=c_W+c_R-n$ and $a \in \{0, 1, \cdots, c-1\}$. We state our result next in Theorem \ref{theorem: stronger general impossibility}. 
 \begin{theorem}
	\label{theorem: stronger general impossibility}	
	A $(\mathcal G=(\mathcal N, \mathcal E), c_W, c_R, \nu=2, 2^K, q)$ multi-version code with side information 
	where there exist $(c-a)$ servers, denoted by $l_0, l_1, \cdots, l_{c-a-1} \in \mathcal N$, such that
	$ \big|  \mathcal N \setminus \bigcup_{i \in \{l_0, l_1, \cdots, l_{c-a-1}\}} \mathcal H_i \big| \geq a $ must satisfy 
	\begin{align}
	\label{general lower bound}
	\log q \geq \min \left\lbrace \frac{1}{c-a}, \frac{2}{c+a} \right\rbrace K,
	\end{align}
where $c=c_W+c_R-n$ and $a \in \{0, 1, \cdots, c-1 \}$.	
\end{theorem}
\noindent We provide the proof of Theorem \ref{theorem: stronger general impossibility} in the Appendix. We next explain briefly the main idea of the proof. 
	\begin{figure}[h]
	\centering
	\includegraphics[width=.9\textwidth,height=.26\textheight]{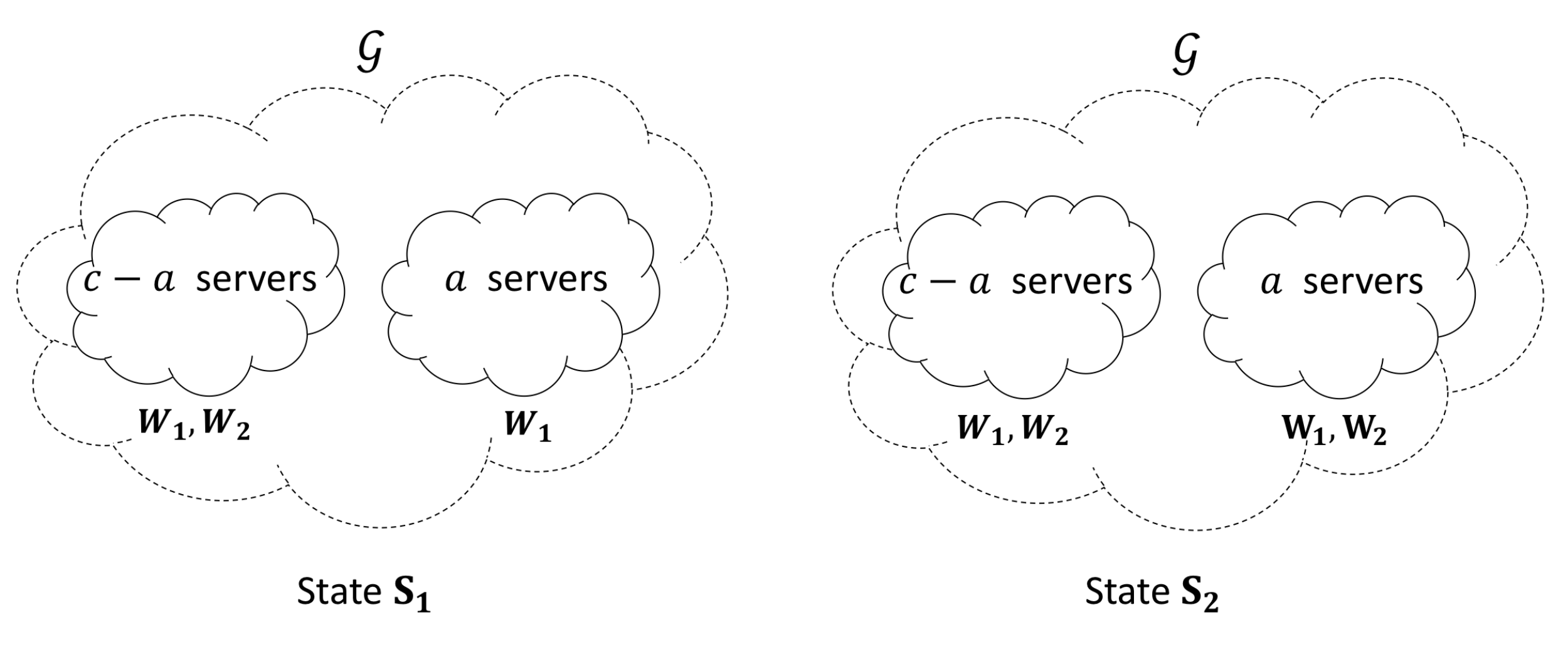}
	\caption{Two system states $\mathbf S_1$ and $\mathbf S_2$ are shown. In $\mathbf S_1$, the decoder can return either $\mathbf W_1$ or $\mathbf W_2$. In $\mathbf S_2$, the decoder must return $\mathbf S_2$. \label{converse1}}
\end{figure}

The main idea of the proof follows by constructing two states $\mathbf S_1 \in \mathcal P([\nu])^n$ and $\mathbf S_2\in \mathcal P([\nu])^n$, such that $L_{\mathbf S_1}=1$ and $L_{\mathbf S_2}=2$ as shown in Fig. \ref{converse1}. Since there exist $(c-a)$ servers, denoted by $l_0, l_1, \cdots, l_{c-a-1} \in \mathcal N$, such that $ \big|  \mathcal N \setminus \bigcup_{i \in \{l_0, l_1, \cdots, l_{c-a-1}\}} \mathcal H_i \big| \geq a $, then there exist $a$ servers denoted by $\{i_0, i_1, \cdots, i_{a-1}\} \subseteq  \mathcal N \setminus \bigcup_{i \in \{l_0, l_1, \cdots, l_{c-a-1}\}} \mathcal H_i $. \\ We construct the two states such that servers $l_0, l_1, \cdots, l_{c-a-1}$ have both versions in the two states and only servers $i_0, i_1, \cdots, i_{a-1}$ change their states from having only $\mathbf W_1$ in $\mathbf S_1$ to having both versions in $\mathbf S_2$. In both states, the decoder connects to the same set of servers denoted by $\mathcal R$, where 
$\mathcal A_{\mathbf S_1}(1) \cap \mathcal R=\mathcal A_{\mathbf S_2}(2) \cap \mathcal R=\{l_0, l_1, \cdots, l_{c-a-1} \} \cup \{i_0, i_1, \cdots, i_{a-1} \}$. Importantly, the servers $l_0, l_1, \cdots, l_{c-a-1}$ cannot differentiate between the two states as they do not know the states of the servers $i_0, i_1, \cdots, i_{a-1}$. \\
In $\mathbf S_2$, $\mathbf W_2$ must be decoded as it is the latest complete version. In $\mathbf S_1$, the decoder can return either $\mathbf W_1$ or $\mathbf W_1$ as $\mathbf W_1$ is the latest complete version. Decoding $\mathbf W_2$ in $\mathbf S_1$ implies that  
\begin{align*}
(c-a) \log q \geq K.
\end{align*}
Decoding $\mathbf W_1$ in $\mathbf S_1$ implies that both versions must be recoverable from the $c$ symbols of servers $\{l_0, \cdots, l_{c-a-1} \} \cup \{i_0, \cdots, i_{a-1} \}$ in $\mathbf S_1$ and the $a$ symbols of servers $\{i_0, \cdots, i_{a-1} \}$ in  $\mathbf S_2$, thus
\begin{align*}
(c+a) \log q \geq 2K.
\end{align*}
Since the decoder in this state can either decode $\mathbf W_1$ or $\mathbf W_2$, we get the lower bound given by Theorem \ref{theorem: stronger general impossibility}.

\begin{remark}
For a side information graph $\mathcal G=(\mathcal N, \mathcal E)$ where the set $A$ containing the possible values of $a$ has multiple values, we have 
\begin{align}
\log q \geq \max_{a \in A} \ \min \left\lbrace \frac{1}{c-a}, \frac{2}{c+a} \right\rbrace K.
\end{align}
\end{remark}

\begin{corollary}
	\label{corollary: stronger impossibility}	
		A $(\mathcal G=(\mathcal N, \mathcal E), c_W, c_R, \nu=2, 2^K, q)$ multi-version code with side information 
	where there exist two servers, denoted by $l_0, l_1 \in \mathcal N$, such that
	$ \big|  \mathcal N \setminus \bigcup_{i \in \{l_0, l_1\}} \mathcal H_i \big| \geq 1 $ must satisfy 	
	\begin{align*}
	\log q \geq K/2,
	\end{align*}
	where $c=c_W+c_R-n=3$.
\end{corollary} \noindent 
The proof of Corollary \ref{corollary: stronger impossibility} follows directly from Theorem \ref{theorem: stronger general impossibility}.

The implication of Corollary \ref{corollary: stronger impossibility} is that, for $c=3$, if two servers are not aware of the state of one server, then the side information does not help in reducing the worst-case storage cost. That follows since a worst-case storage cost of $K/2$ can be achieved in a distributed manner with no side information using the code construction proposed in \cite{wang2018multi}. 
 \subsection{Lower bound for a multi-hop topology}
 We next consider a multi-hop network topology, where the servers are distributed in a ring such that every server is aware of the states of its $h$-hop neighbors as shown in Fig. \ref{multi-hop}. 
 	\begin{figure}[h]
 	\centering
 	\includegraphics[width=.9\textwidth,height=.35 \textheight]{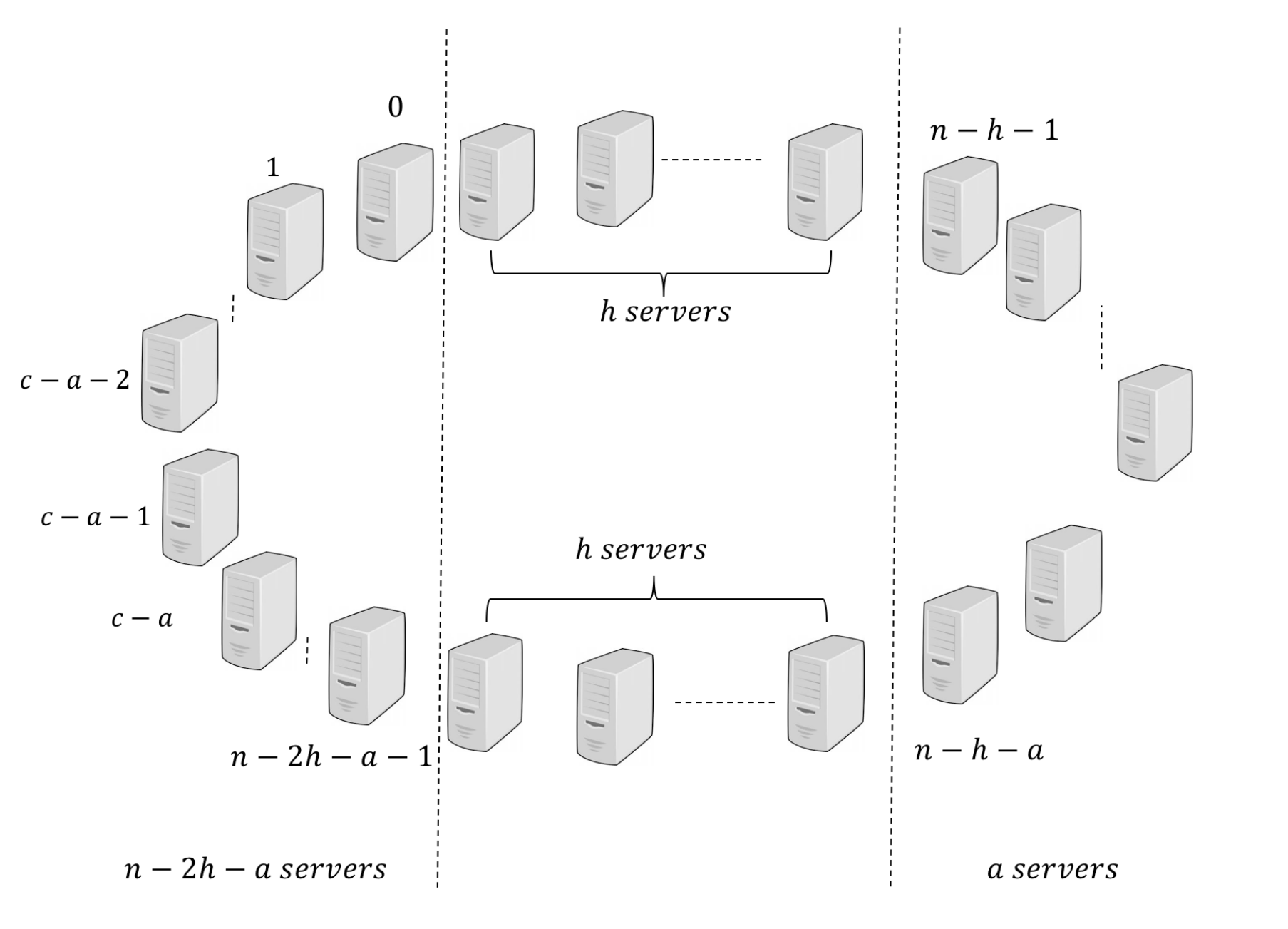}
 	\caption{The multi-hop topology under consideration where the edges between a node and the $h$-hop neighbors are not shown. \label{multi-hop}} 
 \end{figure}
  In this topology, we denote the set of the servers by $\mathcal N=\{0, 1, \cdots, n-1\}$ and the neighborhood of server $i$ is given by $\mathcal H_i=\{i-h, \cdots ,i-2,i-1, i, i+1, i+2, \cdots, i+h\},$ where the addition is modulo $n$. For this topology, Theorem \ref{multi-hop Theorem: stronger impossibility} provides an explicit lower bound on the storage cost.
 \begin{theorem}
 	\label{multi-hop Theorem: stronger impossibility}	
 	A $(\mathcal G=(\mathcal N, \mathcal E), c_W, c_R, \nu=2, 2^K, q)$ multi-version code with side information on the multi-hop topology must satisfy 
 	\begin{align}
 	\log q \geq  \frac{2}{2c-\min(n-2h,c)+\lceil \min(n-2h,c)/3 \rceil} K,
 	\end{align}
where $c=c_W+c_R-n$. 	
 \end{theorem}
\noindent The proof of Theorem \ref{multi-hop Theorem: stronger impossibility} follows also by constructing two states such that some servers cannot differentiate between the two states. For the case where $(n-2h) \geq c$, we have 
\begin{align}
|  \mathcal N \setminus \bigcup_{i \in \{0, 1, \cdots, c-a-1 \}} \mathcal H_i |
&=|\{n-h-a, n-h-a+1, \cdots, n-h-1\}|=a. 
\end{align}
Therefore, we can can apply the result obtained in Theorem \ref{theorem: stronger general impossibility} directly in this case. For the case where $(n-2h) < c$, however, the result of Theorem \ref{theorem: stronger general impossibility} does not apply. We present the detailed proof that handles both cases in the Appendix.
\begin{remark}
For $(n-2h) \leq c$, the lower bound is given by 
\begin{align}
\log q \geq \left(  \frac{1}{c}+\frac{2(n-2h)}{3c^2}+ o\left( \frac{1}{c^3}\right)  \right) K,
\end{align}
and the achievable storage cost of Theorem \ref{first construction theorem} is given by
\begin{align}
\alpha=\left( \frac{1}{c}+\frac{\min{((n-c_W+1)(n-2h-1), c_W-1)}}{c^2} \right)  K.
\end{align}

\end{remark}

\begin{corollary}
\label{corollary: stronger impossibility multi-hop}	
A $(\mathcal G=(\mathcal N, \mathcal E), c_W, c_R, \nu=2, 2^K, q)$ multi-version code with side information on the multi-hop topology must satisfy 
\begin{align}
\log q \geq K/2,
\end{align}
where $h \leq \lfloor (n-3)/2 \rfloor$ and $c=c_W+c_R-n=3$. 	
\end{corollary} 
The proof of Corollary \ref{corollary: stronger impossibility multi-hop} follows directly from Theorem \ref{multi-hop Theorem: stronger impossibility}. 
\noindent Corollary \ref{corollary: stronger impossibility multi-hop} implies that for $c=3$ even if the server is aware of the states of $(n-3)$ other servers, the side information does not reduce the storage cost beyond the case where there is no side information. 
\color{black}
\section{Examples}
\label{Examples}
In this section, we provide numerical examples showing the storage gain of our code constructions. We begin in Section \ref{Numerical Examples} by showing the storage gain for different regimes of the side information. In Section \ref{Amazon Examples}, we show the potential utility of applying our code constructions to Amazon web services (AWS). 
\subsection{Numerical Examples}
\label{Numerical Examples}
In this subsection, we show the storage gain of our achievable schemes as compared with the storage cost of the case where there is no side information. We start with the case where we only have two versions in Example \ref{Two versions Example}.
\begin{example}[Multi-hop network Topology with Two Versions] 
\label{Two versions Example}	
Consider a multi-hop network topology where the servers are distributed in a ring such that every server is aware of the states of its $h$-hop neighbors. In this topology, we denote the set of the servers by $\mathcal N=\{0, 1, \cdots, n-1\}$ and the neighborhood of server $i$ is given by $\mathcal H_i=\{i-h, \cdots ,i-2,i-1, i, i+1, i+2, \cdots, i+h\},$ where the addition is modulo $n$. The achievable storage cost in this case is given by
\begin{align*}
 \alpha=\min \left( {\frac{2}{c+1}, \frac{c+\overline m_{\mathcal G}(c_W)}{c^2}} \right) .
\end{align*}
In Fig. \ref{Two Versions Example}, we show the ratio between the storage cost of Construction \ref{first construction} and the storage cost lower bound in the completely decentralized case $\alpha_0=\frac{2}{c+1} K$ for the case where $n=21$ and $c_W=c_R=19$.
	\begin{figure}[h]
		\centering
		\includegraphics[width=.6\textwidth,height=.24\textheight]{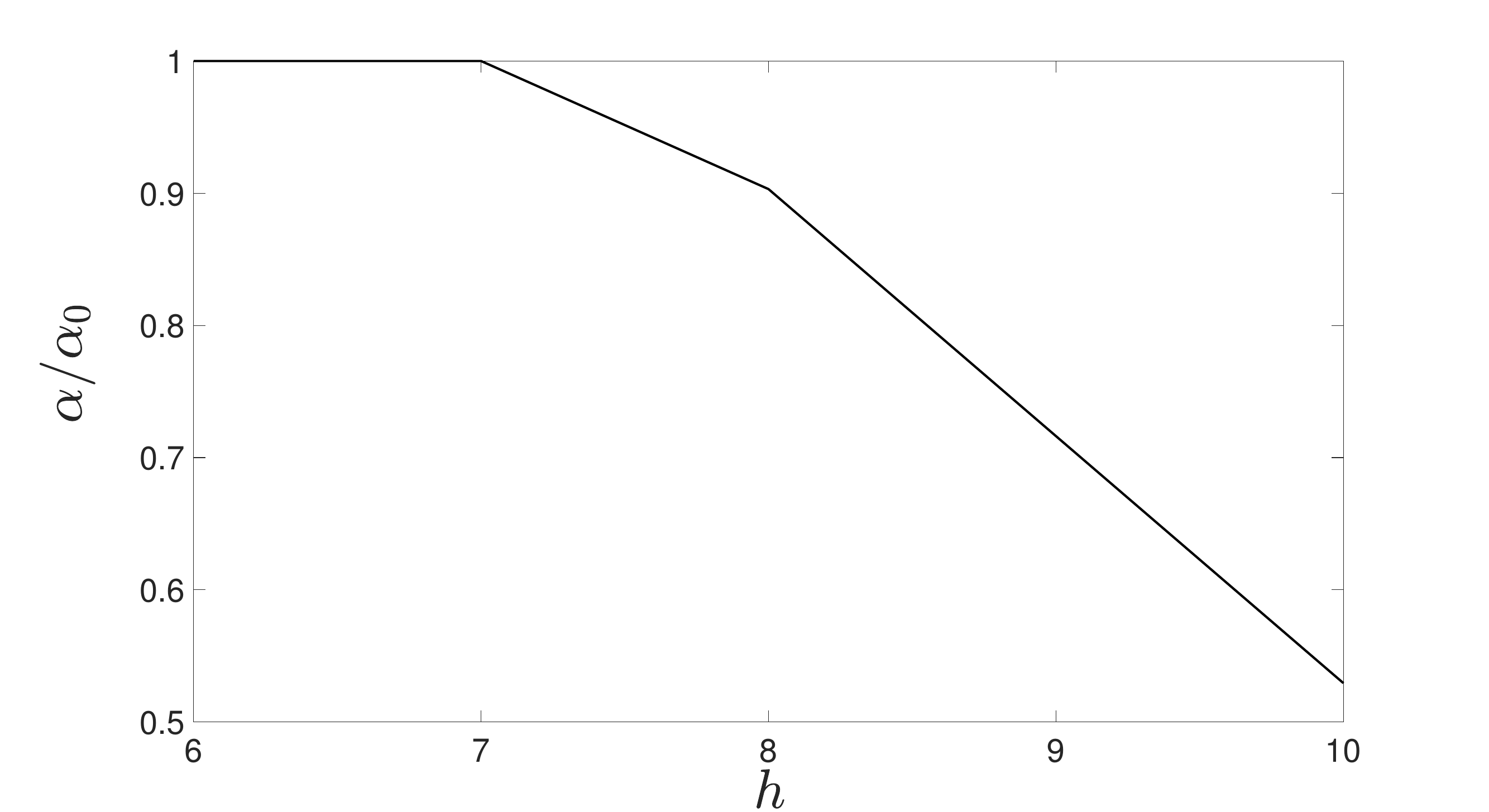}
		\caption{The ratio between the storage cost of Construction \ref{first construction} and storage cost lower bound in the case where there is no side information $\frac{2K}{c+1}$ for $n=21$. \label{Two Versions Example}}
	\end{figure}
\end{example}
\noindent In Example \ref{Three versions Example}, we show the storage gain of our scheme for the case where we have three versions.  
\begin{example}[Multi-hop network Topology with Three Versions]
	\label{Three versions Example}
	Consider a multi-hop storage system where the servers are distributed in a ring and the system tolerates two failures. Assume that $\nu=3$, $n=25$, $c_W=c_R=23$, hence $c=21$. 
	Fig. \ref{Three Versions Example} shows the ratio between the achievable storage cost and the storage cost lower bound in the completely decentralized case $\alpha_0=\frac{\nu}{c+\nu-1}K$ for $\nu=3$.
	\begin{figure}[h]
		\centering
		\includegraphics[width=.6\textwidth,height=.24\textheight]{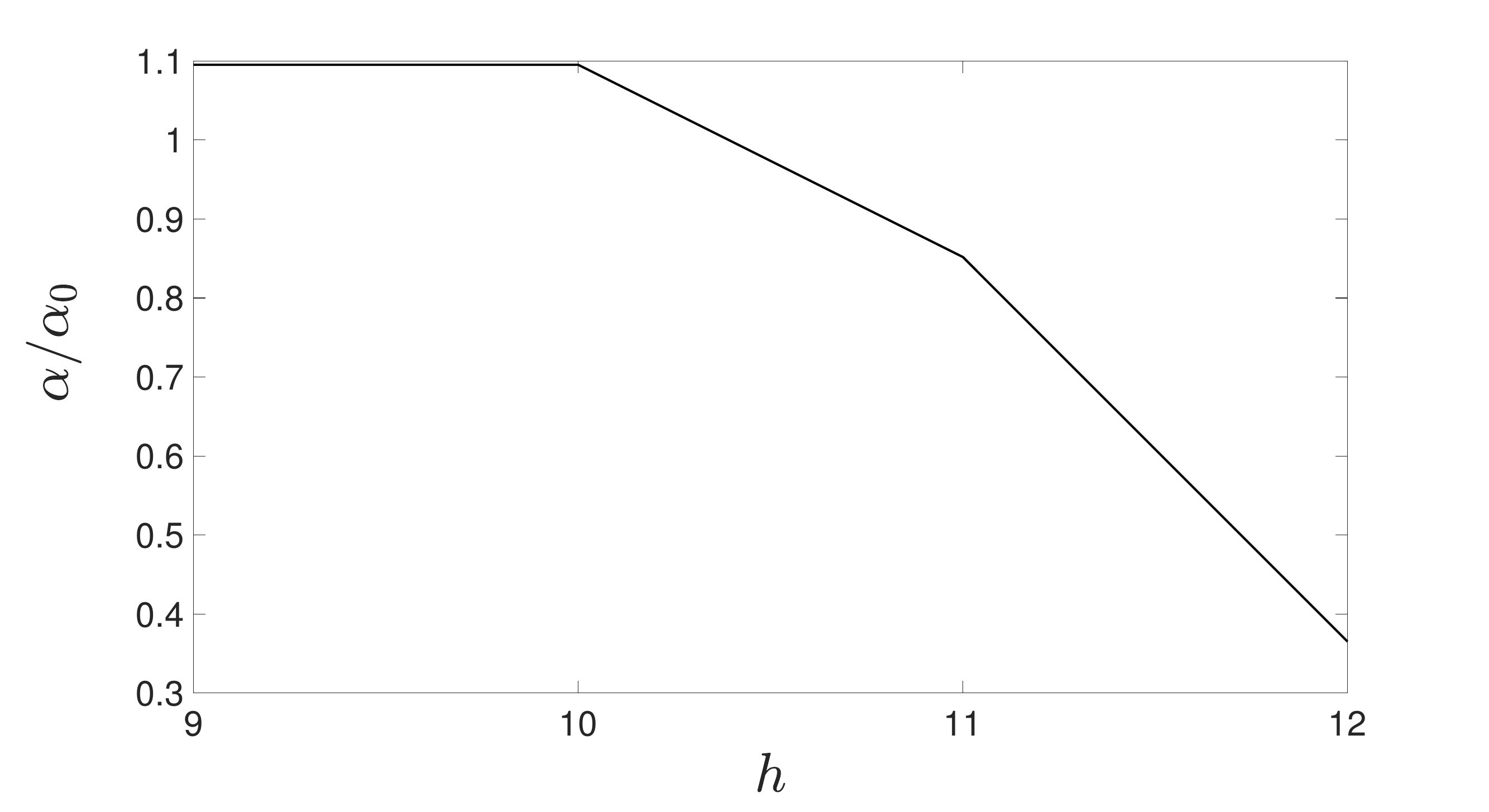}
		\caption{The ratio between the achievable storage cost and the storage cost lower bound in the case where there is no side information $\frac{3K}{c+2}$ for $n=25$. \label{Three Versions Example}}
	\end{figure}
\end{example}
\subsection{Case Study: Amazon Web Services}
\label{Amazon Examples}
	\begin{table*}[h]
	\renewcommand{\arraystretch}{1}
	\centering
	\begin{tabular}{ | p{2cm} | p{2 cm}  | p{2cm} | p{2 cm} | p{2cm} | p{2 cm} | }
		\hline
		Data center & Location & Data center & Location & Data center & Location \\ \hline
		1 & Tokyo &        6& Frankfurt   & 11  &Ohio \\ \hline
		2 & Seoul &         7&  Ireland &12 & N. California  \\ \hline 
		3 & Mumbai & 	 8&London        &13 &Oregon   \\ \hline
		4& Singapore &   9 &  Paris    & &    \\ \hline
		5&  Canada &         10 &  N. Virginia  & &    \\ \hline
	\end{tabular}
	\caption{Data centers locations.}\label{table: locations}
\end{table*}

\begin{table*}[h]
	\renewcommand{\arraystretch}{1}
	\centering
	\begin{tabular}{ | p{2cm} | p{2 cm}  | p{2cm} | p{2 cm} | p{2cm} | p{2 cm} | }
		\hline
		Data center & Storage cost & Data center & Cost & Data center & Storage cost \\ \hline
		1 &  0.019  &          6&    0.0135    & 11 & 0.0125  \\ \hline
		2 &  0.018 &          7&     0.0125 &12 &  0.019  \\ \hline 
		3 &  0.019 & 	        8&   0.0131 &13 &  0.01 \\ \hline
		4&  0.016  &           9 &   0.0131 &&    \\ \hline
		5&   0.0138 &         10 &   0.0125  &&     \\ \hline
	\end{tabular}
	\caption{Storage prices in $\$ / \mathrm{GB}$.}.\label{table: storage prices}
\end{table*}

\begin{table*}[h]
	\renewcommand{\arraystretch}{1}
	\centering
	\begin{tabular}{ | p{0.78 cm} | p{0.6 cm} | p{0.6 cm} |  p{0.6 cm} | p{0.6 cm} | p{0.6 cm} | p{0.6 cm} | p{0.6cm} | p{0.6 cm} | p{0.6 cm} | p{0.6cm} | p{0.6 cm} | p{0.6 cm} | p{0.6 cm} |}
		\hline
		Data center &1 &	2	& 3	& 4	& 5	& 6	& 7	& 8	& 9	& 10 & 11 & 12 & 13  \\ \hline
		1 & 0 & 37.8	&157.2 &90.8 &177.2	&249.7	&234.4	&259.4	&259.4 &167.5	&166.2	&119.6	&106.5	\\ \hline
		2 & 37.9	& 0&	160.1 &	105.7  &	199.7 &	269.9 &	255.7 &	269.3 &	268.2  &	190.7 &	189.3 &	153 &	128.2 	\\ \hline
		3 &136.9  & 181.5 &0 &68.8  &212.8 &129.9 &134.4 & 128 &118.3  & 187.7 & 202.2 &240.8 &225 \\ \hline
		4  & 90 &112.4 &82.3 &0  &240.9 &189.7 &186.4 &181.3 &178.5 &267.8 &232.6 &184.7 &194.7 \\ \hline
		5 & 159.2 &189.5 &202 &222.3 &0&103.1 &81.7 &92 &95.4 &17.8 &27.2 &82 &81.7 \\ \hline
		6 & 241.3 & 267.3 &115.3 &174.8  &107 &0&24.2 &19.1 &12.8  &90.4 &98.9 &147.8 &165.4 \\ \hline
		7  & 230 &258.4 &128.4 &180 &85.2 &23.8 &0&14.6 &21.6  &72.7 &84.6 &152.8 &137.4 \\ \hline
		8 & 236.9 &265.3 &116.9 &168  &93.9 &15.7 &13.2 &0&10.7  &78 &88.7 &141.7 &148.5 \\ \hline
		9 & 233.5 &301.6 &111.6 &173 &97.6 &14.4 &20.4 &11 &0  &81.7 &99.4 &140.7 &157.8 \\ \hline
		10& 164.3 &188.8 &195.8 &239.9  &18.8 &92 &73.1 &79.8 &110.5 &0 &13.66 &67.2 &79.3 \\ \hline
		11& 162.4 &189.9 &199.7 &226  &27.6 &121.5 &87.7 &91.3 &94.6  &16.4 &0&55.9 &74.53 \\ \hline
		12& 111.4 &157.9 &253.4 &178.3  &81.7 &148.7 &150.7 &140 &146.7  &67.8 &53.9 &0&23.4 \\ \hline
		13& 109.8 &139.7 &226 &166.5  &73.4 &167.8 &137.8 &150.8 &160.4  &84 &73 &25.8&0
		\\ \hline
	\end{tabular}
	\caption{Latency between data centers in $\mathrm{ms}$.}\label{table: latency}
\end{table*}
In this subsection, we show the potential utility of applying our schemes to the data centers of Amazon. Table \ref{table: locations} and Table \ref{table: storage prices}  provide the data centers locations and the storage prices obtained from \cite{prices} as of 03/11/2019. In Table \ref{table: latency}, we provide the latency between $13$ data center of Amazon obtained from \cite{latency} as of 03/11/2019. We assume that a shared object is stored over the data centers $\mathcal N=\{1, 2, 3, 4, 5, 6, 7, 8, 9, 10, 11, 12, 13 \}$, $\nu=2$ and $c_W=c_R=11$, thus $c=9$. 

We note that exchanging the side information comes with a latency increase for write operations, as the data centers need to wait to hear from other data centers about the versions they received before deciding what to store. By bounding the maximum allowable latency increase, we can obtain the side information graph $\mathcal G=(\mathcal N, \mathcal E)$. Specifically, an edge exists from a data center $i \in \mathcal N$ to a data center $j \in \mathcal N$, if the latency from the $j$-th data center to the $i$-th data center is below a pre-specified latency bound. Here, we use the latencies based on Table \ref{table: latency} to explore the role of our code constructions, via a trade-off between the storage costs and latencies as shown in Fig. \ref{Storage-latency}. Intuitively, allowing higher latency, may add more edges to the side information graph $\mathcal G$ and that can decrease $\overline m_{\mathcal G}(c_W)$ which improves the storage cost. 

\noindent Note that our storage costs and latencies are \emph{projected} values based on Table \ref{table: latency}. The reported latencies should be interpreted as the additional latency incurred by write operations for writing a new value that would be incurred due to the exchange of side information. The actual latency can be higher as protocols (e.g., \cite{ABD, CadambeCoded_NCA, Dutta}) have additional rounds of communication, for instance, to discern the latest logical timestamp, etc.

\begin{figure}[h]
		\centering
\includegraphics[width=.6\textwidth,height=.24\textheight]{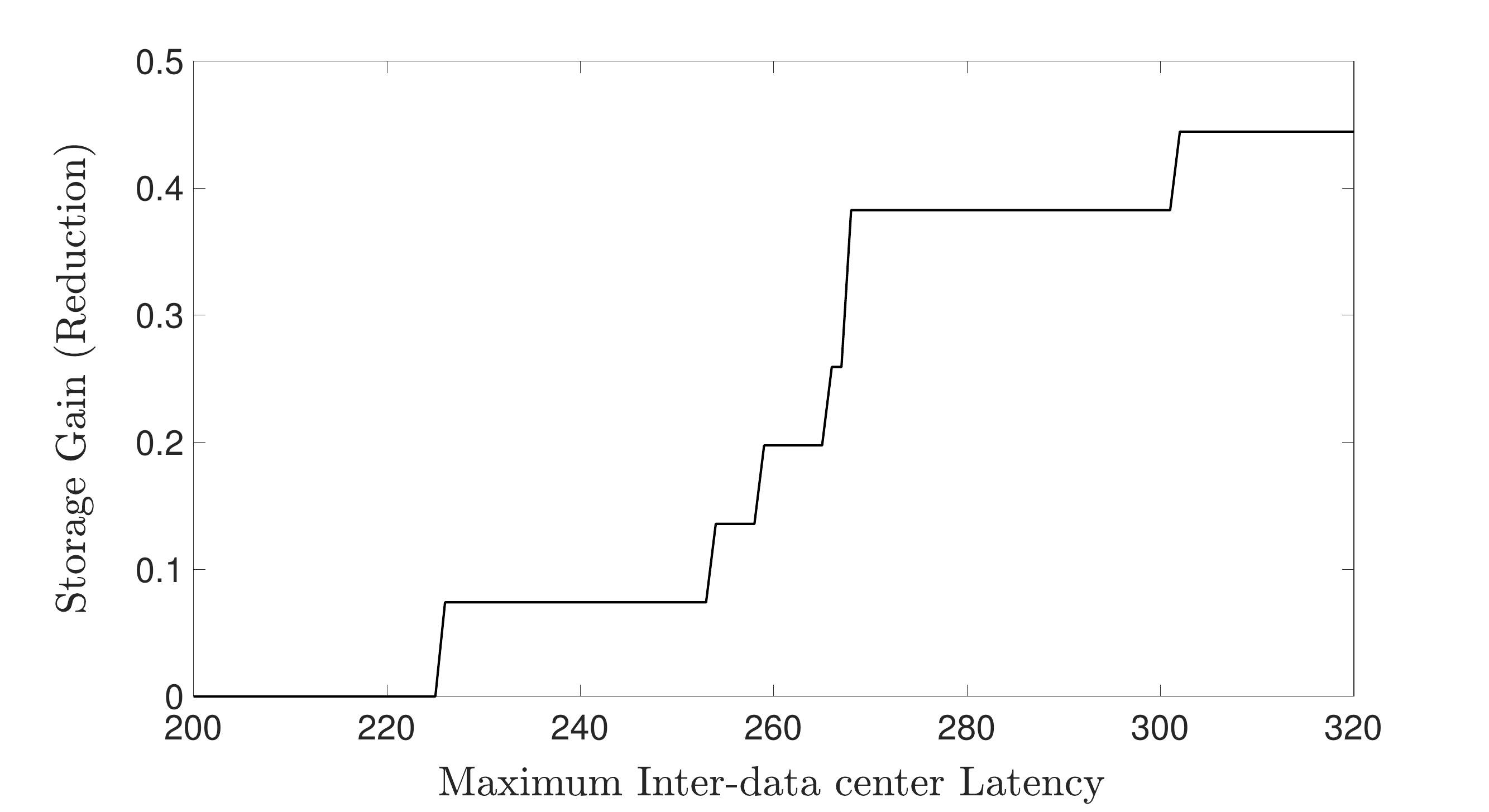}
\caption{Trade-off between the projected storage cost and the projected latency increase. \label{Storage-latency}}
\end{figure}

We now compare between the no side information case, the partial side information case and the complete side information case in terms of the additional latency due to side information exchange and the corresponding storage cost reductions. 
\begin{enumerate}[label=\Alph*.]
\item No side information
\begin{itemize}
	\item The inter-data center latency increase due to the side information exchange is equal to $0$.
	\item The per-server storage cost in this case is given by
\begin{align*}
	\alpha=\frac{2}{c+1}K= 0.2K.
\end{align*}
The storage prices in dollars of the system is given in Table \ref{table: storage prices}. We denote the storage price of the $i$-th data center by $p_i$, where $i \in \mathcal N$. The total storage price denoted by $P$ is expressed as follows
\begin{align*}
P= \alpha \sum_{i \in \mathcal N} p_i,
\end{align*}
and in our case $P=\$ 0.0384 \ K $.
\end{itemize}
\item   Partial side information
\begin{itemize}

	\item \textbf{Latency Increase:} In the case where the data centers do not exchange their states (completely decentralized system), the servers do not wait before deciding what to store. In the partial side information case, the servers wait for the side information of the other servers to be received before deciding what to store. Therefore, there is a latency increase in the partial side information case as compared with the case of no side information. For the sake of example, we assume that the maximum allowable inter-data center latency increase due to side information exchange is $260 \ \mathrm{ms}$.  
	
\item The per-server storage cost is given by
	\begin{align*}
	\alpha=\frac{c+\overline m_{\mathcal G}(c_W)}{c^2}K.
	\end{align*}
While a side information graph that corresponds to the complete side information case has $169$ edges - every edge between the $13$ data centers including self edges - the side information graph $\mathcal G$ that corresponds to maximum allowable latency increase of $260 \ \mathrm{ms}$ has only $162$ edges. For instance, there is no edge between the Seoul data center (\#2) and Frankfurt Data center (\#6) in this case. In this partial side information case, using a brute-force computer search, we can verify that
	\begin{align*}
	\overline m_{\mathcal G}(c_W) =  \max_{\mathbf S \in \mathcal P([\nu])^n: \mathcal A_{\mathbf S}(u) = (c_W-1)} \ |\{i \in \mathcal A_{\mathbf S}(u): |\mathcal A_{\mathbf S}(u) \cap \mathcal H_i| \geq c_W+|\mathcal H_i|-n \}|=4.
	\end{align*}
	Therefore, the per-server storage cost is given by $$\alpha=0.16 K$$ and the total storage price is equal to $$ P= \alpha \sum_{i \in \mathcal N} p_i=\$ 0.0307
	 \ K.$$
	\textbf{Storage Gain:} The lower bound on the storage cost in case of no side information implies a storage cost that is at least $0.2K-o(K)$ bits. Therefore, in the partial side information case, we get a storage gain of $19.75 \%$ of our achievable scheme as compared with the optimal achievable scheme with no side information.
\end{itemize}
\item Complete side information
\begin{itemize}
	\item The inter-data center latency increase due to side information exchange has to be at least $301.6 \ \mathrm{ms}$ to allow all data centers to exchange their states. Thus, there is a latency increase of at least $301.6 \ \mathrm{ms}$ as compared with the no side information case. 
	\item The per-server storage cost in bits in this case is given by 
	\begin{align*}
	\alpha=\frac{1}{c} K=0.1111K.
	\end{align*}
The total storage price	in this case is given by
$$P= \alpha \sum_{i \in \mathcal N} p_i= \$      0.0213
 K.$$
\end{itemize}
\end{enumerate}
We summarize the additional latency-storage trade-off in Table \ref{table: comparison}.
	
\begin{table*}[h]
	\centering	
	\begin{tabular}{ | c | c | c|}
		\hline
		Case & Maximum inter-data center latency &  Storage price  \\ \hline      
		No side information& $0$ &  $ \$0.0384 \ K$ \\ [1ex]\hline
		Partial side information & $260 \ \mathrm{ms}$  
		 & $\$    0.0307
		 \ K$  \\ [1ex]\hline 
		Complete side information & $ 301.6 \ \mathrm{ms}$ 
		& $\$     0.0213
		 \ K$  \\ [1ex]\hline 
	\end{tabular}
	\caption{The maximum inter-data center latencies (i.e., latency increase due to side information exchange) and the corresponding storage costs for various regimes.}.\label{table: comparison}
\end{table*}

\section{Conclusion}  
\label{Conclusion}
 In this paper, we have studied geo-distributed key-value stores where a data center can acquire side information of the data versions propagated to some other servers or data centers based on the underlying topology. We have provided code constructions showing that the exchanging this side information results in a better storage cost compared to the no side information for some regimes at the expense of the higher latency 
 of exchanging the side information gossip messages. We have also demonstrated the potential cost reductions and latency-storage trade-off of our constructions through a case study over Amazon web services. Interestingly, our converse results identify topologies where exchanging side information gossip messages does not improve the storage cost. Designing a protocol based on our code constructions is an interesting area of future research.  
 \newpage
\appendices
\label{Appendix}
\section{Proof of Theorem \ref{theorem: stronger general impossibility}	}
\begin{proof}
	We construct two states $\mathbf S_1$ and $\mathbf S_2$ with different decoding requirements such that the set of servers $\{l_0, l_1, \cdots, l_{c-a-1}\}$ cannot differentiate between the two states due to the limited side information.  In particular, $\mathbf S_1$ and $\mathbf S_2$ are constructed such that only $a$ servers, denoted by $\{i_0, i_1, \cdots, i_{a-1} \} \subseteq \mathcal N \setminus  \bigcup_{i \in \{l_0, l_1, \cdots, l_{c-a-1}\}} \mathcal H_i $ change their states from $\mathbf S_1$ to $\mathbf S_2$.
	\begin{enumerate}
		\item \textbf{State $\mathbf S_2$}. In this state, $\mathbf W_2$ is the latest complete version and hence it must be decoded. In particular, $c_W$ servers have both $\mathbf W_1$ and $\mathbf W_2$ and the remaining $n-c_W$ do not have any version. The set of servers that have $\mathbf W_{[2]}$ is given by the following disjoint union 
		\begin{align*}
		\mathcal A_{\mathbf S_2}(1)=\mathcal A_{\mathbf S_2}(2)=\{l_0, l_1, \cdots, l_{c-a-1}\} \cup \{l_{c-a}, l_{c-a+1}, \cdots, l_{c_W-a-1}\} \cup \{i_0, i_1, \cdots, i_{a-1}\},
		\end{align*}
		where $\{l_{c-a}, l_{c-a+1}, \cdots, l_{c_W-a-1}\} \subset \mathcal N$.
	In this state, the decoder connects to the following set of servers 
		\begin{align*}
		\mathcal R=\mathcal N \setminus \{l_{c-a}, l_{c-a+1}, \cdots, l_{c_W-a-1}\}.
		\end{align*}
		We denote the value stored at the $i$-th server, $i \in \mathcal N$, in this state by $ X_i =\varphi_{\mathbf S_2(\mathcal H_i)}^{(i)} (\mathbf W_{\mathbf S_2(i)}) \in \mathcal [q]. $
		Since $\mathbf W_2$ is the latest complete version in this state, we must have 
		\begin{align}
		H(\mathbf W_2| X_{\{l_0, l_1, \cdots, l_{c-a-1} \} \cup \{i_0, i_1, \cdots, i_{a-1} \}})=0.
		\end{align}
		Thus, we have the following inequalities 
		\begin{align}
		\label{state S2 bound}
		\sum_{j \in \{l_0, \cdots, l_{c-a-1}\} \cup \{i_0, \cdots, i_{a-1}\} } H(X_j| \mathbf W_1) &\geq H(X_{\{l_0, l_1, \cdots, l_{c-a-1} \} \cup \{i_0, i_1, \cdots, i_{a-1} \}}|\mathbf W_1) \notag \\  
		&= H(\mathbf W_2| \mathbf W_1)-H(\mathbf W_2| \mathbf W_1, X_{\{l_0, l_1, \cdots, l_{c-a-1} \} \cup \{i_0, i_1, \cdots, i_{a-1} \}}) \notag \\
		&\stackrel{(a)}{=}H(\mathbf W_2)-H(\mathbf W_2| X_{\{l_0, l_1, \cdots, l_{c-a-1} \} \cup \{i_0, i_1, \cdots, i_{a-1} \}}) \notag \\
		&=K,
		\end{align}
		where $(a)$ follows since we assume that $\mathbf W_1, \mathbf W_2$ are independent and uniformly distributed over $[2^K]$ as the code should work for any distribution of $\mathbf W_{[2]}$.

		\item \textbf{State $\mathbf S_1$}. In this state,  $\mathbf W_1$ is the latest complete version. Therefore, the decoder can either return $\mathbf W_2$ or $\mathbf W_1$. In particular, $c_W-a$ servers have both $\mathbf W_1$, $\mathbf W_2$, $a$ servers have only $\mathbf W_1$ and the remaining servers do not have any version. The set of servers that have $\mathbf W_1$ is given by 
		\begin{align*}
		\mathcal A_{\mathbf S_1}(1)=\{l_0, \cdots, l_{c-a-1}\} \cup \{l_{c-a}, \cdots, l_{c_W-a-1} \} \cup \{i_0, \cdots, i_{a-1}\},
		\end{align*} 
		and the set of servers that have $\mathbf W_2$ is given by
		\begin{align*}
		\mathcal A_{\mathbf S_1}(2)=\{l_0, \cdots, l_{c-a-1}\} \cup \{l_{c-a}, \cdots, l_{c_W-a-1} \}. 
		\end{align*}	 
		Suppose that the decoder connects to the following set of servers 
		\begin{align*}
		\mathcal R=\mathcal N \setminus \{l_{c-a}, l_{c-a+1}, \cdots, l_{c_W-a-1}\}.
		\end{align*}
		We denote the value stored at the $i$-th server, $i \in \mathcal N$, in this state by $Y_i = \varphi_{\mathbf S_1(\mathcal H_i)}^{(i)} (\mathbf W_{\mathbf S_1(i)}) \in [q]$. Since servers $l_0, l_1, \cdots, l_{c-a-1}$ observe the same information in both states, we have $Y_i=X_i$, for $i \in \{l_0, l_1, \cdots, l_{c-a-1} \}$.
		In this state, the decoder must either return $\mathbf W_2$ or $\mathbf W_1$. We consider these cases next. 
		\begin{enumerate}
			\item In order to decode $\mathbf W_2$ in this state, we must have 
			\begin{align}
			H(\mathbf W_2| X_{\{l_0, l_1, \cdots, l_{c-a-1} \}})=0.
			\end{align}
			Consequently, we have the following inequities 
			\begin{align*}
			(c-a) \log q & \stackrel{(a)}{\geq} \sum_{j \in \{l_0, l_1, \cdots, l_{c-a-1}\}} H(X_j) \notag \\
			&\geq  \sum_{j \in \{l_0, l_1, \cdots, l_{c-a-1}\}} H(X_j| \mathbf W_1) \notag \\
			& \geq H(X_{\{l_0, l_1, \cdots, l_{c-a-1} \}} | \mathbf W_1)
			\notag \\
			& = H(X_{\{l_0, l_1, \cdots, l_{c-a-1} \}} | \mathbf W_1)-H(X_{\{l_0, l_1, \cdots, l_{c-a-1} \}} | \mathbf W_1, \mathbf W_2) \notag \\
			&= H(\mathbf W_2| \mathbf W_1)- H(\mathbf W_2| \mathbf W_1, X_{\{l_0, l_1, \cdots, l_{c-a-1} \}}) \notag \\
			&=K,
			\end{align*}
			where $(a)$ follows since $X_i \in [q], \forall i \in \mathcal N$. Therefore, the storage cost is lower-bounded as follows in this case 
			\begin{align}
			\log q \geq \frac{1}{c-a} K.
			\end{align}
			\item In order to decode $\mathbf W_1$ in this state, we must have 	
			\begin{align}
			H(\mathbf W_1| X_{\{l_0, l_1, \cdots, l_{c-a-1} \}}, 
			Y_{\{i_0, i_1, \cdots, i_a\}})=0.
			\end{align}
			Consequently, we have the following inequalities 
			\begin{align}
			\label{decoding version 1}
			& \sum_{j \in \{l_0, \cdots, l_{c-a-1}\}} H(X_j | \mathbf W_2)+\sum_{j \in \{i_0, \cdots, i_{a-1}\}} H(Y_j | \mathbf W_2) \notag \\ &\geq H(X_{\{l_0, l_1, \cdots, l_{c-a-1} \}}, 
			Y_{\{i_0, i_1, \cdots, i_a\}}| \mathbf W_2)
			\notag \\ &= H(X_{\{l_0, l_1, \cdots, l_{c-a-1} \}}, 
			Y_{\{i_0, i_1, \cdots, i_a\}}| \mathbf W_2)-H(X_{\{l_0, l_1, \cdots, l_{c-a-1} \}}, 
			Y_{\{i_0, i_1, \cdots, i_a\}}| \mathbf W_{[2]}) \notag \\
			&= H(\mathbf W_1| \mathbf W_2)-H(\mathbf W_1| \mathbf W_2, X_{\{l_0, l_1, \cdots, l_{c-a-1} \}}, 
			Y_{\{i_0, i_1, \cdots, i_a\}}) \notag \\
			&=K. 
			\end{align}
			Moreover, since $H(X_i| \mathbf W_{[2]})=0, \ i \in \mathcal N$ and  $\mathbf W_1$ and $\mathbf W_2$ are independent, we have
			\begin{align*}
			& H(X_i)=H(X_i|\mathbf W_1)+H(X_i|\mathbf W_2), \notag \\
		&	H(Y_i)=H(Y_i|\mathbf W_1)+H(Y_i|\mathbf W_2),
			\end{align*}
			$\forall i \in \mathcal N$. Therefore, (\ref{decoding version 1}) can be rewritten as follows
			\begin{align*}
			\sum_{j \in \{l_0, \cdots, l_{c-a-1}\}} H(X_j)+\sum_{j \in \{i_0, \cdots, i_{a-1}\}} H(Y_j) &\geq K+\sum_{j \in \{l_0, \cdots, l_{c-a-1}\}} H(X_j| \mathbf W_1) \notag \\ &+\sum_{j \in \{i_0, \cdots, i_{a-1}\}} H(Y_j| \mathbf W_1) \notag \\
			&= K+\sum_{j \in \{l_0, \cdots, l_{c-a-1}\}} H(X_j| \mathbf W_1),
			\end{align*}
			where the last equality follows since $H(Y_j| \mathbf W_1)=0, \forall j \in \{i_0, \cdots, i_{a-1}\}$ as those servers only have $\mathbf W_1$. This implies with (\ref{state S2 bound}) the following
			\begin{align*}
			c \log q &\geq \sum_{j \in \{l_0, \cdots, l_{c-a-1}\}} H(X_j)+\sum_{j \in \{i_0, \cdots, i_{a-1}\}} H(Y_j) \notag \\ 
			&\geq K+\sum_{j \in \{l_0, \cdots, l_{c-a-1}\}} H(X_j| \mathbf W_1) \notag \\
			&\geq 2K-\sum_{j \in \{i_0, \cdots, i_{a-1}\}} H(X_j| \mathbf W_1).
			\end{align*}
			Therefore, the storage cost in this case is lower-bounded as follows 
			\begin{align}
			\log q \geq \frac{2}{c+a}K.
			\end{align}
		\end{enumerate}
		Since in state $\mathbf S_1$ the decoder can decode either $\mathbf W_1$ or $\mathbf W_2$, the storage cost is lower-bounded as follows
		\begin{align}
		\log q \geq  \min \left\lbrace \frac{K}{c-a}, \frac{2K}{c+a} \right\rbrace. 
		\end{align} 
	\end{enumerate}
	
\end{proof}

\section{Proof of Theorem \ref{multi-hop Theorem: stronger impossibility}}
\begin{proof}
	We consider the multi-hop network and construct two states $\mathbf S_2$ and $\mathbf S_1$. The two states have different decoding requirements, but the set of servers $\{0, 1, \cdots, n-2h-a-1\}$, where $a \in \{0, 1, \cdots, \min(n-2h, c)-1 \}$, cannot differentiate between the two states. To keep the notation simple, we denote $\min(n-2h, c)$ by $t$.  
	\begin{enumerate}
		\item   \textbf{State $\mathbf S_2$}. In this state $\mathbf W_2$ is the latest complete version. Therefore, the decoder must return $\mathbf W_2$ in this state. The
		set of servers that have $\mathbf W_1$ is the same as the set of servers that have $\mathbf W_2$ and is given by the following disjoint union
		\begin{align*}
		\mathcal A_{\mathbf S_2}(1) &=\mathcal A_{\mathbf S_2}(2)=\{0, 1, \cdots, t-a-1\} \cup \{l_{t-a}, l_{t-a+1}, \cdots, l_{c_W-a-1} \} \\ &\cup \{n-h-a, \cdots, n-h-1\},
		\end{align*}
		where $\{l_{t-a}, l_{t-a+1}, \cdots, l_{c_W-a-1} \} \subset \mathcal N$.
		The decoder connects to the following set of servers to decode $\mathbf W_2$
		\begin{align*}
		\mathcal R=\mathcal N \setminus \{l_{t-a}, l_{t-a+1}, \cdots, l_{t+c_W-c-a-1}\}.
		\end{align*}
		The symbol stored by the $i$-th server in this state is denoted by $X_i =\varphi_{\mathbf S_2(\mathcal H_i)}^{(i)} (\mathbf W_{\mathbf S_2(i)})$, where $i \in \mathcal N$. 
		Since $\mathbf W_2$ is the latest complete version in this state, we must have 
		\begin{align}
		H(\mathbf W_2| X_{ \{0, 1, \cdots, t-a-1 \} \cup \{l_{t+c_W-c-a}, \cdots, l_{c_W-a-1} \} \cup \{n-h-a, \cdots, n-h-1 \}})=0.
		\end{align}
		Therefore, we have the following inequalities 
		\begin{align}
		\label{multi-hop state S2 bound}
		&\sum_{j \in \{0, 1, \cdots, t-a-1\} \cup \{l_{t+c_W-c-a}, \cdots, l_{c_W-a-1} \} \cup \{n-h-a, \cdots, n-h-1\}} H(X_j| \mathbf W_1) \notag \\ & \geq H(X_{ \{0, 1, \cdots, t-a-1 \} \cup \{l_{t+c_W-c-a}, \cdots, l_{c_W-a-1} \} \cup \{n-h-a, \cdots, n-h-1 \}}| \mathbf W_1)\notag \\ & - H(X_{ \{0, 1, \cdots, t-a-1 \} \cup \{l_{t+c_W-c-a}, \cdots, l_{c_W-a-1} \} \cup \{n-h-a, \cdots, n-h-1 \}}| \mathbf W_{[2]}) \notag \\ 
		&=H(\mathbf W_2| \mathbf W_1)-H(\mathbf W_2| \mathbf W_1, X_{ \{0, 1, \cdots, t-a-1 \} \cup \{l_{t+c_W-c-a}, \cdots, l_{c_W-a-1} \} \cup \{n-h-a, \cdots, n-h-1 \}}) \notag \\ 
		&=K.
		\end{align}
		\item   \textbf{State $\mathbf S_1$}. In this state $\mathbf W_1$ is the latest complete version. The decoder in this case must either return $\mathbf W_1$ or $\mathbf W_2$. The set of servers that have $\mathbf W_1$ is given by 
		\begin{align*}
		\mathcal A_{\mathbf S_1}(1)=\{0, 1, \cdots, t-a-1\} \cup \{l_{t-a}, l_{t-a+1}, \cdots, l_{c_W-a-1} \} \cup \{n-h-a, \cdots, n-h-1\}.
		\end{align*}
		The set of servers that have $\mathbf W_2$ is given by
		\begin{align*}
		\mathcal A_{\mathbf S_1}(2)=\{0, 1, \cdots, t-a-1\} \cup \{l_{t-a}, l_{t-a+1}, \cdots, l_{c_W-a-1} \}.
		\end{align*} 
		The decoder connects to the following set of servers to decode either $\mathbf W_1$ or $\mathbf W_2$ 
		\begin{align*}
		\mathcal R=\mathcal N \setminus \{l_{t-a}, l_{t-a+1}, \cdots, l_{t+c_W-c-a-1}\}.
		\end{align*}	
		The symbol stored at the $i$-th server in this state is denoted by $Y_i = \varphi_{\mathbf S_1(\mathcal H_i)}^{(i)} (\mathbf W_{\mathbf S_1(i)})$, where $i \in \mathcal N$. Since servers $0, 1, \cdots, n-2h-a$ observe the same information in both states, we have $Y_i=X_i$, for $i \in \{0, 1, \cdots, n-2h-a \}$. \\ Depending on which version the decoder will return, we consider the following cases. 
		\begin{enumerate}
			\item In order to decode $\mathbf W_2$ in this state, we must have  
			\begin{align}
			H(\mathbf W_2| X_{ \{0, 1, \cdots, t-a-1 \}}, Y_{ \{l_{t+c_W-c-a}, \cdots, l_{c_W-a-1} \}})=0.
			\end{align}
			Consequently, we have the following inequities 
			\begin{align*}
			(c-a) \log q  &\stackrel{(a)}{\geq}  \sum_{j \in \{0,1, \cdots, t-a-1\}} H(X_j)+\sum_{j \in \{l_{t+c_W-c-a}, \cdots, l_{c_W-a-1}\}} H(Y_j) \notag \\
			& \geq \sum_{j \in \{0,1, \cdots, t-a-1\}} H(X_j| \mathbf W_1)+\sum_{j \in \{l_{t+c_W-c-a}, \cdots, l_{c_W-a-1}\}} H(Y_j| \mathbf W_1) \notag \\
			& \geq H(X_{ \{0, 1, \cdots, t-a-1 \}}, Y_{ \{l_{t+c_W-c-a}, \cdots, l_{c_W-a-1} \}}|\mathbf W_1) \notag \\	
			& = H(X_{ \{0, 1, \cdots, t-a-1 \}}, Y_{ \{l_{t+c_W-c-a}, \cdots, l_{c_W-a-1} \}}|\mathbf W_1)\notag \\ &-H(X_{ \{0, 1, \cdots, t-a-1 \}}, Y_{ \{l_{t+c_W-c-a}, \cdots, l_{c_W-a-1} \}}|\mathbf W_{[2]}) \notag \\
			&=H(\mathbf W_2| \mathbf W_1)-H(\mathbf W_2| \mathbf W_1, X_{ \{0, 1, \cdots, t-a-1 \}}, Y_{ \{l_{t+c_W-c-a}, \cdots, l_{c_W-a-1} \}}) \notag \\
			&=K,
			\end{align*}
			where $(a)$ follows since $X_i \in [q]$ and $Y_i \in [q], \forall i \in \mathcal N$. Therefore, decoding $\mathbf W_2$ in this state implies a storage cost that is lower-bounded as follows
			\begin{align}
			\log q \geq \frac{1}{c-a}K.
			\end{align}

			\item In order to decode $\mathbf W_1$ in this state, we must have
			\begin{align*}
			H(\mathbf W_1| X_{ \{0, 1, \cdots, t-a-1 \}}, 
			Y_{ \{l_{t+c_W-c-a}, \cdots, l_{c_W-a-1} \} \cup \{n-h-a, \cdots, n-h-1 \}})=0.
			\end{align*}
			Consequently, we have the following inequalities 
			\begin{align}
			&\sum_{j \in \{0, 1, \cdots, t-a-1\}} H(X_j|\mathbf W_2)+\sum_{j \in \{l_{t+c_W-c-a}, \cdots, l_{c_W-a-1}\}} H(Y_j|\mathbf W_2)+\sum_{j \in \{n-h-a, \cdots, n-h-1\}} H(Y_j|\mathbf W_2) \notag \\ &\geq H( X_{ \{0, 1, \cdots, t-a-1 \}}, 
			Y_{ \{l_{t+c_W-c-a}, \cdots, l_{c_W-a-1} \} }, Y_{ \{n-h-a, \cdots, n-h-1 \}}| \mathbf W_2) \notag \\ &- H(X_{ \{0, 1, \cdots, t-a-1 \}}, 
			Y_{ \{l_{t+c_W-c-a}, \cdots, l_{c_W-a-1} \} }, Y_{ \{n-h-a, \cdots, n-h-1 \}}| \mathbf W_{[2]}) \notag 
 \\
			&=H(\mathbf W_1| \mathbf W_2)-H(\mathbf W_1| \mathbf W_2, X_{ \{0, 1, \cdots, t-a-1 \}}, 
			Y_{ \{l_{t+c_W-c-a}, \cdots, l_{c_W-a-1} \} }, Y_{ \{n-h-a, \cdots, n-h-1 \}}) \notag \\
			&=K.
			\end{align}
			Since $\mathbf W_1$ and $\mathbf W_2$ are independent, we also have 
			\begin{align}
			H(X_i)&=H(X_i|\mathbf W_1)+H(X_i|\mathbf W_2) \notag \\
			H(Y_i)&=H(Y_i|\mathbf W_1)+H(Y_i|\mathbf W_2),
			\end{align}
			$\forall i \in \mathcal N$. Therefore, we have 
			\begin{align*}
			c \log q &\geq \sum_{j \in \{0, 1, \cdots, t-a-1\}} H(X_j)+\sum_{j \in \{l_{t+c_W-c-a}, \cdots, l_{c_W-a-1}\}} H(Y_j)+\sum_{j \in \{n-h-a, \cdots, n-h-1\}} H(Y_j) \notag \\
			& \geq K+\sum_{j \in \{0, 1, \cdots, t-a-1\}} H(X_j|\mathbf W_1)+\sum_{j \in \{l_{t+c_W-c-a}, \cdots, l_{c_W-a-1}\}} H(Y_j| \mathbf W_1)\\ &+\sum_{j \in \{n-h-a, \cdots, n-h-1\}} H(Y_j| \mathbf W_1) \notag \\
			& = K+\sum_{j \in \{0, 1, \cdots, t-a-1\}} H(X_j|\mathbf W_1)+\sum_{j \in \{l_{t+c_W-c-a}, \cdots, l_{c_W-a-1}\}} H(Y_j| \mathbf W_1),
			\end{align*}
			where the last equality follows as $H(Y_j| \mathbf W_1)=0, \forall j \in \{n-h-a, \cdots, n-h-1\}$ since those servers only have received $\mathbf W_1$.  This implies with (\ref{multi-hop state S2 bound}) the following 
			\begin{align*}
			c \log q &\geq 2K-\sum_{j \in \{l_{t+c_W-c-a}, \cdots, l_{c_W-a-1}\} \cup \{n-h-a, \cdots, n-h-1\}} H(X_j|\mathbf W_1) \\ &+ \sum_{j \in \{l_{t+c_W-c-a}, \cdots, l_{c_W-a-1}\}} H(Y_j| \mathbf W_1).
			\end{align*}
			Therefore decoding $\mathbf W_1$ implies a storage cost that lower-bounded as follows
			\begin{align}
			\log q \geq \frac{2}{2c-\min(c, n-2h)+a}K.
			\end{align}
		\end{enumerate}
		Since in state $\mathbf S_1$ the decoder can decode either $\mathbf W_1$ or $\mathbf W_2$, the storage cost is lower-bounded as follows
		\begin{align}
		\log q \geq  \min \left\lbrace \frac{1}{c-a}, \frac{2}{2c-\min(c, n-2h)+a} \right\rbrace K. 
		\end{align} 
		Choosing $a=\lceil \frac{\min(n-2h, c)}{3}\rceil$, we get 
		\begin{align}
		\log q \geq \frac{2}{2c-\min(n-2h, c)+\lceil \min(n-2h, c)/3\rceil} K.
		\end{align}
	\end{enumerate}
\end{proof}
\newpage
\bibliographystyle
{IEEEtran}
\bibliography{IEEEabrv,Nulls}

\end{document}